\documentclass[sigconf, nonacm]{acmart}

\usepackage[linesnumbered,lined,boxed,commentsnumbered, ruled, noend]{algorithm2e}
\usepackage{algorithmic}
\usepackage{booktabs}
\usepackage{amsmath,amsfonts,amsthm}
\usepackage{dsfont}
\usepackage{footmisc}
\usepackage{graphicx}
\usepackage{hyperref}
\usepackage[utf8]{inputenc}
\usepackage{listings}
\usepackage{makecell}
\usepackage{multirow}
\usepackage{subcaption}
\usepackage{tablefootnote}
\usepackage{tabularx}
\usepackage{textcomp}
\usepackage{tikz}
\usepackage{xcolor}

\usetikzlibrary{
arrows %
,arrows.meta%
,automata%
,backgrounds %
,bending%
,calc%
,decorations.markings %
,decorations.pathmorphing %
,decorations.pathreplacing%
,decorations.text %
,fit %
,patterns%
,pgfplots.groupplots%
,positioning %
,shadings %
,shadows %
,shadows.blur%
,shapes%
,shapes.callouts %
,shapes.geometric %
,shapes.misc %
,shapes.multipart%
,tikzmark%
,trees %
}

\newcommand\vldbdoi{XX.XX/XXX.XX}

\newcommand\vldbavailabilityurl{http://vldb.org/pvldb/format_vol14.html}
\newcommand\vldbpagestyle{plain} 

\newtheorem{definition}{Definition}

\newtheorem{example}{Example}

\newtheorem{theorem}{Theorem}

\newcommand{\pred}[1]{\textsf{\small #1}}

\begin{document}

\title{Evaluating Complex Queries on Streaming Graphs}



\author{Anil Pacaci}

\affiliation{%
  \institution{University of Waterloo}
}

\email{apacaci@uwaterloo.ca}

\author{Angela Bonifati}

\affiliation{%
  \institution{Lyon 1 University}
}

\email{angela.bonifati@univ-lyon1.fr}

\author{M. Tamer \"{O}zs{u}}

\affiliation{%
  \institution{University of Waterloo}
}
\email{tamer.ozsu@uwaterloo.ca}

\begin{abstract}
We study the problem of evaluating persistent queries over streaming graphs in a principled fashion. These queries need to be evaluated over unbounded and very high speed graph streams.
We define a streaming graph data model and query model incorporating navigational queries, subgraph queries and paths as first-class citizens. To support this full-fledged query model we develop a streaming graph algebra that describes the precise semantics of persistent graph queries with their complex constructs. We present  transformation rules and describe query formulation and plan generation for persistent graph queries over streaming graphs. Our implementation of a streaming graph query processor 
shows the feasibility of our approach and allows us to gauge the high performance gains obtained for query processing over streaming graphs.

\end{abstract}

\maketitle
\pagestyle{\vldbpagestyle}


\section{Introduction}
\label{sec:intro}
Many modern applications in various domains now operate on very high speed streaming graphs ~\cite{vldbj19_Sahu:2017aa} .
For example, Twitter's recommendation system ingests 12K events/sec on average \cite{grewal2018recservice}, Alibaba transaction graph processes 30K edges/sec at its peak \cite{qiu2018real}. 
Efficient querying of these streaming graphs is a crucial task for these applications that monitor complex graph patterns and relationships. 

Existing graph DBMSs mostly follow the traditional database paradigm where data is persistent and queries are transient; consequently, they do not support persistent query semantics where queries are registered into the system and results are generated incrementally as the graph edges arrive.
Persistent queries on streaming graphs enable users to continuously obtain new results on rapidly changing data, supporting online analysis and real-time query processing, the latter being an important functionality of future graph processing engines ~\cite{abs-2012-06171}. This is demonstrated in the following example. 

\begin{figure}
\centering
    \includegraphics[width=\linewidth]{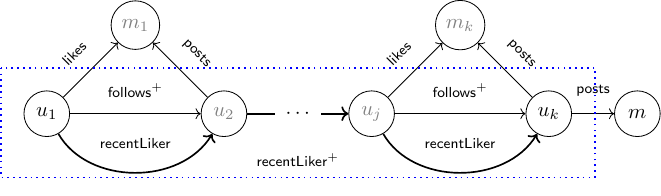}
    \caption{Complex graph pattern representing the query in Exanmple \ref{ex:alert}}
    \label{fig:running_example}
\end{figure}

\begin{example}\label{ex:alert}
In many online social networking applications users  post original content, sometimes link this to other users' content and react to each other's posts.
We say that a user $u_2$ is a {\normalfont \pred{recentLiker}} for another user $u_1$ if $u_2$ has recently liked posts that are created by $u_1$ and $u_2$, and $u_1$ are following each other.
The goal of the recommendation service is to notify users, in real-time, of new content that are posted by others that are connected by a path of {\normalfont \pred{recentLiker}} relationship -- these constraints are modeled as a complex graph pattern as the one shown in Figure \ref{fig:running_example}. 
The service might provide the context for its recommendations by returning the full paths of people who are recent likers such as the path between users $u_1$ and $u_k$.
This real-time notification task is an example of a  persistent query over the streaming graph of user interactions that returns the recommended content in real-time. 
\end{example}

Real-world applications that feature complex graph patterns as the one shown in the above example require: 
\begin{itemize}
    \item 
    \textbf{(R1)} subgraph queries that find matches of a given graph pattern (e.g. in Figure \ref{fig:running_example} the triangle pattern involving \pred{posts}, \pred{likes} and transitive closure of the \pred{follows} relationship)
    \item 
    \textbf{(R2)} path navigation queries that traverse paths based on user specified constraints (e.g. in Figure \ref{fig:running_example} arbitrary-length paths of the \pred{recentLiker} relationship); and
    \item 
    \textbf{(R3)} the ability to treat paths as first-class citizens of the data model, hence to manipulate and return paths (e.g. in Figure \ref{fig:running_example} the query returns the full paths of \pred{recentLiker}).
\end{itemize}
Even in the context of one-time queries over static graphs, these requirements are poorly addressed by existing graph DBMSs and their languages. Existing
query languages (e.g., PGQL, SPARQL v1.1, Cypher) address the first two issues by replacing edge labels of a subgraph pattern with regular expressions -- this is known as \emph{unions of conjunctive RPQs} (UCRPQ) \cite{wood2012query, bonifati2018querying}.
However, UCRPQ lacks algebraic closure and composability, limiting optimization opportunities.
Furthermore, the output of a path navigation query is typically a set of pairs of vertices that are connected by a path under the constraints of a given regular expression.
Hence, UCRPQ-based query languages limit path navigation queries to boolean reachability, without the ability to return and manipulate paths.
G-CORE~\cite{angles2018g} addresses these limitations at the language specification level and has
influenced the standardization efforts for a query language for graph DBMSs (see \url{https://www.gqlstandards.org/}).
To the best of our knowledge, there is no work that uniformly addresses all three requirements. 

The issues become more complex in the context of persistent query processing over streaming graphs, which is the focus of this paper.  
A number of specialized algorithms focus on evaluating subgraph pattern queries on streaming graphs \cite{vldb18_AmmarMSJ18, li2019time, qiu2018real, kim2018turboflux, edbt15_choudhury:2015aa}, and our previous work focuses on path navigation queries and introduces the first streaming algorithms for RPQ evaluation \cite{Pacaci_2020}, a limited subset of UCRPQ.
However, a general-purpose model and framework that address the above discussed requirements of real-world applications (which feature complex graph patterns) in a uniform and principled manner is currently missing.

Querying streaming data in real-time imposes additional and novel requirements: 
\begin{itemize}
    \item 
    \textbf{(R4)} graph streams are unbounded, making it infeasible to employ batch algorithms on the entire stream; and
    \item 
    \textbf{(R5)} graph edges arrive at a very high rate and real-time answers are required as the graph emerges.
\end{itemize}
Existing graph DBMSs based on the \textit{snapshot} model are not able to keep up with the high arrival rates \cite{pacaci:2017aa}. 
The unsuitability of existing graph DBMSs for querying streaming data has motivated the design of specialized systems addressing singular application needs (e.g., \cite{qiu2018real, grewal2018recservice, edbt15_choudhury:2015aa, li2019time, kim2018turboflux}).
The lack of systematic support for query processing over streaming graphs hinders the development of a general-purpose query processor for streaming graphs.

In this paper, we study the design of a general-purpose query processor for streaming graphs that addresses all of the above-discussed requirements in a uniform and principled manner.
In analogy to traditional DBMSs, 
our framework provides the foundational tools to realize the well-known steps of query processing for streaming graph queries as follows:
\begin{enumerate}
    \item a streaming graph query expressed in a declarative, high-level user language is translated into a query plan that consists of logical operators with precise semantics;
    \item algebraic transformation rules are used to explore the plan space through query rewrites to find a ``good'' one, paving the way for query optimization;
    \item the execution plan is built by selecting appropriate physical implementations of logical operators  that are incremental and non-blocking;
    \item the execution engine continuously executes the persistent query upon arrival of new edges to obtain new results.
\end{enumerate}

Our focus in this paper is on the following key aspects of this framework:
\begin{enumerate}
\item We  introduce the formal \textit{Streaming Graph Query} model (SGQ -- Section \ref{sec:query-system}) based on a streaming graph model (Section \ref{sec:data-model}). SGQ is based on the Regular Query (RQ) model\footnote{A computationally well-behaved subset of Datalog with tractable query evaluation and decidable query containment, similar to relational algebra \cite{reutter2017regular}.}, and  provide precise semantics for persistent graph queries with subgraph patterns, path navigations and windowing constructs. Most of all, SGQ treats  paths as first-class citizens, enabling queries to return and manipulate paths. Moreover, SGQ is closed under transitive closure and composable. 

\item We present the \textit{Streaming Graph Algebra} (SGA -- Section \ref{sec:algebra}) as a foundational basis for evaluating SGQ and 
provide an algorithm for translating SGQ into SGA expressions (Section \ref{sec:query-formulation}).
Utilizing SGA's closedness (Section \ref{sec:algebra-composability}), we 
introduce transformation rules for novel SGA operators for systematic exploration of the plan space (Section \ref{sec:query-transformation}).

\item We describe query execution plans (Section \ref{sec:query-processor}) and non-blocking physical operator implementations (Section \ref{sec:physical-algebra}).

\item We describe a prototype implementation of a streaming graph query processor based on  \textit{Timely Dataflow}~\cite{murray2013naiad}, i.e., the dataflow computational model (Section \ref{sec:implementation}).
\end{enumerate}

Unlike existing work on streaming graphs that relies on ad hoc algorithms, our algebraic approach provides the foundational framework to precisely describe the semantics of complex persistent graph queries over streaming graphs and to optimize such queries by query optimizers.
SGA unifies path navigation and subgraph pattern queries in a structured manner (\textbf{R1} \& \textbf{R2}), i.e., it attains composability by properly closing UCRPQ under recursion.
In addition, our framework treats paths as first-class citizens of its data model, enabling the proposed algebra to express queries that return and manipulate paths (\textbf{R3}).
Extensive experimental analysis over real and synthetic datasets  of our prototype implementation that incorporates incremental, non-blocking algorithms as physical operators (\textbf{R4}  \& \textbf{R5}) demonstrates the feasibility and the  performance gains of our approach.
To the best of our knowledge, this is the first work to study the design of a streaming graph query processor, and to introduce an end-to-end solution to evaluating streaming graph queries with complex constructs.

\label{sec:related}
\section{Related Work}

\subsection{Stream Processing Systems}

Early research on stream processing primarily adapt the relational model and its query operators to the streaming setting
(e.g., STREAM ~\cite{abw02}, Aurora ~\cite{abadi2003}, Borealis ~\cite{cidr05_abadiabcchlmrrtxz05}).
In contrast, modern Data Stream Processing Systems (DSPS) such as Storm ~\cite{toshniwal2014storm}, Heron \cite{sigmod15_kulkarni:2015}, Flink \cite{carbonekemht15}
do not necessarily offer a full set of DBMS functionality.
Existing literature  (as surveyed by Hirzel et al. \cite{HirzelBBVSV18}) heavily focus on general-purpose systems and do not consider core graph querying functionality such as \textit{subgraph pattern matching} and \textit{path navigation}.

Existing work on streaming graph systems, by and large, target graph analytics workloads with little or no focus on graph querying functionalities on which we focus in this paper (e.g., STINGER ~\cite{ediger2012stinger}, GraphOne ~\cite{kumar2019graphone, kumar2020graphone},  GraphIn ~\cite{sengupta2016graphin},  GraphTau ~\cite{iyer2016time}, GraPu ~\cite{sheng2018grapu}, GraphBolt ~\cite{mariappan2019graphbolt}).
The primary focus of these systems is to build and maintain a sequence of snapshots for iterative graph analytics workloads under a stream of updates.
SPARQL extensions have also been proposed for persistent query processing over RDF streams (e.g.,  C-SPARQL \cite{barbieri2009c}, CQELS \cite{le2011native}, SPARQL$_{stream}$~\cite{calbimonte2010enabling} and RSP-QL \cite{dell2015towards}). These are the most similar to our setting, but they are designed for SPARQLv1.0 and cannot formulate path expressions such as RPQs that cover more than 99\% of all recursive queries found in massive Wikidata query logs {\cite{bonifati2019navigating}}.
Also, query processors of these systems do not employ incremental operators, except 
Sparkwave~\cite{komazec2012sparkwave} that focuses on stream reasoning.
Our proposed framework supports complex graph patterns arising in existing graph query languages, including SPARQL v1.1 property paths, and introduces non-blocking operators optimized for streaming workloads.

\subsection{Incremental View Maintenance}

A persistent query over sliding windows can be formulated as an IVM problem, where the view definition is the query itself and window movements are updates to the underlying database.
Although conceptually similar, the algorithms (e.g., the classical \textit{Counting} ~\cite{gms93} algorithm) and systems (e.g., DBToaster ~\cite{KochAKNNLS14}, F-IVM ~\cite{nikolic2018incremental}, ViewDF ~\cite{yang:2017aa}) in this category primarily target non-recursive queries.
General-purpose IVM techniques for 
recursive queries (such as \textit{DRed}~\cite{gms93} and  \textit{Absorption Provenance}~\cite{icde09_4812481}) ignore the structure of graph queries and the inherent temporal patterns of streaming graphs, resulting in significant computational overhead to compute and store a large number of derivations.
Differential Dataflow (DD) ~\cite{cidr13_mcsherrymii13} is a state-of-the-art distributed system for incremental maintenance of possibly cyclic dataflows.
Graphsurge ~\cite{graphsurge2020} uses DD as the underlying execution layer to share computation across multiple, possibly recursive views over static graphs.
As we show in this paper (Section \ref{sec:physical-algebra}), DD can also be used for evaluating SGQ.
Our framework, in contrast to these, exploits the structure of graph queries and the temporal patterns of sliding windows to minimize the cost of evaluating complex queries on streaming graphs (Section \ref{sec:experiments_comparative}).

\subsection{Graph Algorithms}
TriAL~\cite{libkin2018trial} and Temporal Graph Algebra (TGA) \cite{moffitt2017temporal} adopt an algebraic approach to graph query processing, similar to ours.
TriAL is designed to be used for one-time navigational queries over static triplestores, and it cannot be used as a standalone graph query language.
In contrast, our proposed SGA is designed as a standalone language for persistent graph queries over streaming graphs, and it can express complex graph patterns expressed in high-level user languages (Section \ref{sec:gcore}).
TGA extends temporal relational operators to PGM, and its Spark implementation introduces physical operator implementations for its algebraic primitives \cite{aghasadeghi2020zooming}.
However, it is designed for exploratory graph analytics over the entire history of changes.
In contrast, our framework is designed to continuously evaluate graph queries as the underlying (potentially unbounded) streaming graph changes.


The closest to ours are \textit{specialized} algorithms on dynamic and streaming graphs \cite{vldb18_AmmarMSJ18, li2019time, qiu2018real, kim2018turboflux, edbt15_choudhury:2015aa, Pacaci_2020}.
Some of these \cite{kim2018turboflux, edbt15_choudhury:2015aa} study the incremental evaluation of subgraph pattern queries.
Their focus is developing efficient incremental algorithms to maintain matches of a given subgraph pattern as the underlying graph changes.
Ammar et al.  \cite{vldb18_AmmarMSJ18} present distributed worst-case-optimal join algorithms for subgraph pattern queries.
Li et al.  \cite{li2019time} study subgraph isomorphism search over streaming graphs with timing-order constraints.
GraphS~\cite{qiu2018real} introduces efficient index structures that are optimized for cycle detection queries.
In previous work \cite{Pacaci_2020}, we study the design space of algorithms for path navigation over streaming graphs and provide algorithms for persistent evaluation of RPQ, a  subset of the class of queries that we address in this paper.
In contrast we target a \textit{general} query model that captures the
precise semantics of streaming graph queries.

\section{Data Model: Streaming Graphs}
\label{sec:data-model}

\subsection{Preliminaries}
\label{sec:preliminaries}

\begin{definition}[Graph]\label{def:graph}
A directed labeled graph is a quin\-tuple $G = (V, E, \Sigma, \psi, \phi)$ where $V$ is a set of vertices, 
$E$ is a set of edges, 
$\Sigma$ is a set of labels,
$\psi: E \rightarrow V \times V$ is an incidence function
and $\phi: E \rightarrow \Sigma$ is an edge labelling function. 
\end{definition}

\begin{definition}[Path and Path Label]\label{def:path}
\sloppy Given $u, v \in V$, a path $p$ from $u$ to $v$ in graph $G$ is a sequence of edges $ u \stackrel{p}{\rightarrow} v : \langle e_1, \cdots, e_n \rangle$.
The label sequence of a path $p$ is defined as the concatenation of edge labels, i.e., $\phi^p(p) = \phi(e_1) \cdots \phi(e_n) \in \Sigma^*$.
\end{definition}

We use  $\mathbb{T} = (\mathcal{T}, \leq)$  to define a discrete, total ordered time domain and use timestamps $t \in \mathcal{T}$ to denote time instants.  
Without loss of generality, the rest of the paper uses non-negative integers to represent timestamps.

\begin{definition}[Streaming Graph Edge] \label{def:streaming-edge}
A \textit{streaming graph edge} (sge) is a quadruple $(src, trg, l, t)$ where $src, trg \in V$ are endpoints of an edge $e \in E$, $l \in \Sigma$ is the label of the edge $e$, and $t \in \mathcal{T}$ is the event (application) timestamp assigned by the source, i.e., $\psi(e) = (src, trg)$ and $\phi(e) = l$. \footnote{We assume that sges are generated by a single source and arrive in order, and leave out-of-order arrival as future work. }
\end{definition}

\begin{definition}[Input Graph Stream]\label{def:inputstream}
An input graph stream is a continuously growing sequence of streaming graph edges $S^I = \big[sge_1, sge_2, \cdots \big]$ where each sge represents an edge in graph $G$ and sges are non-decreasingly ordered by their timestamps.\footnote{We use $\big[\big]$ to denote ordered streams throughout the paper}
\end{definition}

Input graph streams represent external sources that provide the system with the graph-structured data. 
Our proposed framework uses a different format that generalizes Definition \ref{def:inputstream} to also represent intermediate results and outputs of persistent queries (Definition \ref{def:insertstreamtuple}). 

\begin{definition}[Validity Interval]
A \textit{validity interval} is a half-open time interval $[ts, exp)$ consisting of all distinct time instants $t \in \mathcal{T}$ for which $ts \leq t < exp$.
\end{definition}

Timestamps are commonly used to represent the time instant in which the interaction represented by the sge occured \cite{qiu2018real, Pacaci_2020, li2019time}.
Alternatively, we use intervals to represent the period of validity of sges.
In this paper, we argue that using \textit{validity intervals} leads to a succinct representation and simplifies operator semantics by separating the specification of window constructs from operator implementation.
As an example, each sge with timestamp $t$ can be assigned a validity interval $[t, t+1)$ that corresponds to a single time unit with smallest granularity that cannot be decomposed into smaller time units.\footnote{Commonly referred as \texttt{NOW} windows as described in Section \ref{sec:algebra}.}
Similarly, an sge $e=(u, v, l, [ts, exp))$ with a validity interval is equivalent to a set of sges $\{ (u, v, l, t_1), \cdots, (u, v, l, t_n) \}$ where $t_1 = ts$ and $t_n = exp - 1$.
Windowing operator (to be precisely defined momentarily in Section \ref{sec:algebra}) are used to assign validity intervals based on the windowing specifications of a given query.

\subsection{Streaming Graphs}
\label{sec:streaming-graph-model}

We now describe the logical representation of streaming graphs that is used throughout the paper.
First, we extend the directed labeled graph model with materialized paths to represent paths as first-class citizens of the data model. 
As per Definition \ref{def:path}, a path between vertices $u$ and $v$ is a sequence of edges $ u \stackrel{p}{\rightarrow} v : \langle e_1, \cdots, e_n \rangle$ that connects vertices $u$ and $v$, i.e., the path $p$ defines a higher-order relationship between vertices $u$ and $v$ through a sequence of edges.
By treating paths as first-class citizens like vertices and edges, the materialized path graph model enables queries to have paths as inputs and outputs.
In addition, it enables edges and paths to be stitched together to form complex graph patterns as will be shown in Section \ref{sec:algebra}.

\begin{definition}[Materialized Path Graph]\label{def:path-graph}
A materialized path graph is a 7-tuple $G = (V, E, P, \Sigma, \psi, \rho, \phi)$ where $V$ is a set of vertices, 
$E$ is a set of edges,
$P$ is a set of paths,
$\Sigma$ is a set of labels,
$\psi: E \rightarrow V \times V$ is an incidence function,
$\rho: P \rightarrow E \times \cdots \times E $ is a total function that assigns each path to a finite, ordered sequence of edges in $E$, 
and $\phi: (E \cup P) \rightarrow \Sigma$ is a labeling function, where images of $E$ and $P$ under $\phi$ are disjoint, i.e., $\phi(E) \cap \phi(P) = \emptyset$.
\end{definition}

The function $\rho$ assigns to each $p: u \stackrel{p}{\rightarrow} v \in P$ an actual path $\langle e_1, \cdots, e_n \rangle$ in graph $G$ satisfying: for every $i \in [ 1, \cdots, n)$, $\psi(e_i) = (src_i, trg_i)$, $ trg_{i} = src_{i+1},$ and $src_1 = u, trg_n = v$.
Materialized path graph is a strict generalization of the directed labeled graph model (Definition \ref{def:graph}), i.e., each directed labeled graph $G$ is also a materialized path graph where $P = \emptyset$.
We now generalize the notion of streaming graph edges (Definition \ref{def:streaming-edge}) as follows:

\begin{definition}[Streaming Graph Tuple]\label{def:insertstreamtuple}
A \textit{streaming graph tuple} (sgt) is a quintuple $sgt = (src, trg, l, [ts, exp), \mathcal{D}) $
where distinguished (explicit) attributes $src,trg \in V$  are endpoints of an edge $e \in E$ or a path $p \in P$ in graph $G$ and $l \in \Sigma$ is its label, and non-distinguished (implicit) attributes $[ts, exp) \in \mathcal{T} \times \mathcal{T}$ is a half-open time-interval representing $t$'s validity and $\mathcal{D}$ is a payload consists of edges in $E$ that participated in the generation of the tuple $t$. 
\end{definition}

Streaming graph tuples generalize sges (Definition \ref{def:streaming-edge}) to represent, in addition to input graph edges, derived edges (new edges as operator and query results that are not necessarily part of the input graph) and paths (sequence of edges as operator and query results).
We use the notation $E^I \subset E$ to denote the set of input graph edges, and $\phi(E^I)$ to denote the fixed set of labels that are reserved for input graph edges.
Additionally, non-distinguished (implicit) attribute $\mathcal{D}$ of an sgt $t$ captures the path $p$, i.e., sequence of edges, in case the sgt $t$ represents a path. Otherwise, $\mathcal{D}$ is the edge $e$ that the sgt $t$ represents.

\begin{definition}[Streaming Graph]\label{def:streaming-graph}
 \sloppy A streaming graph $S$ is a continuously growing sequence of streaming graph tuples $S= \big[t_1, t_2, \cdots \big]$ in which each tuple $t_i$ arrives at a particular time $t_i$ ($t_i < t_j$ for $i<j$). 
\end{definition}


\begin{figure}
\centering
    \includegraphics[width=\linewidth]{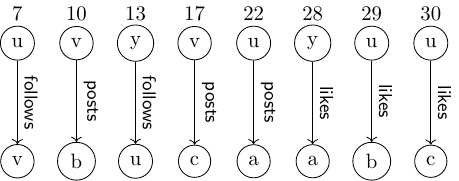}
    \caption{The input graph stream from an online social network of Example \ref{ex:alert}}
    \label{fig:input-stream}
\end{figure}
\begin{figure}
\centering
    \includegraphics[width=\linewidth]{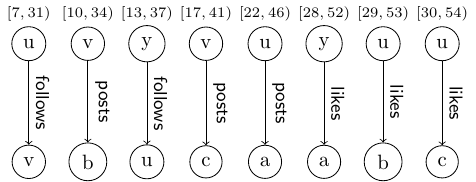}
    \caption{Streaming graph obtained from the input graph stream in Figure \ref{fig:input-stream} where the validity interval of each element is set based on a $24h$ window}
    \label{fig:streaming-graph}
\end{figure}
\begin{figure}
\centering
    \includegraphics[width=\linewidth]{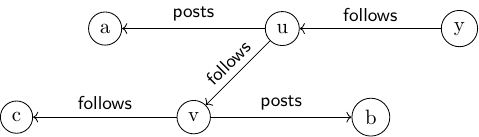}
    \caption{Snapshot graph of of the streaming graph in Figure \ref{fig:streaming-graph} at  $t=25$.}
    \label{fig:snapshot-graph}
\end{figure}

Figure \ref{fig:input-stream} depicts an excerpt of the input graph stream of the application in Example \ref{ex:alert}, where each tuple represent an interaction and each timestamp represent the time instant that the interaction occurs.
Its corresponding streaming graph where each tuple is assigned a time interval is shown in Figure \ref{fig:streaming-graph} (validity intervals are assigned by  \pred{WSCAN} operator -- see Section \ref{sec:algebra}).

Unless otherwise specified, we consider streaming graphs to be \textit{append-only}, i.e., each sgt represents an insertion, and use the \textit{direct approach} to process expirations due to window movements.
Explicit deletions of previously arrived sgts can be supported by explicitly manipulating the validity interval of a previously arrived sgt~\cite{kramer2009semantics}.
This corresponds to the \textit{negative tuple} approach \cite{ghanem2006incremental, golabo05}.
Section \ref{sec:physical-algebra} describes the processing of insertions, deletions and expirations under alternative window semantics for physical operator implementations. 

\begin{definition}[Logical Partitioning]\label{def:logical-partitioning}
A logical partitioning of a streaming graph $S$ is a label-based partitioning of its tuples and it produces a set of disjoint streaming graphs $\{ S_{l_1}, \cdots, S_{l_n} \}$ where each $S_{l_i}$ consists of sgts of $S$ with the label $l_{i}$, i.e., $S = \bigcup_{l \in \Sigma}(S_l)$
\end{definition}

This label-based partitioning of streaming graphs provides a coherent representation for inputs and outputs of operators in logical algebra (Section \ref{sec:algebra}).
At the logical level, it can be performed by the \textit{filter} operator of the logical algebra (precisely defined in Definition \ref{def:selection}), 
and logical operators of our algebra process logically partitioned streaming graphs as their inputs and outputs unless otherwise specified.

\begin{definition}[Value-Equivalence]\label{def:equivalence}
\sloppy
Sgts  $ t_1 = (u_1, v_1, l_1,$  $[ts_1, exp_1), \mathcal{D}_1) $ and $t_2 = (u_2, v_2, l_2, [ts_2, exp_2), \mathcal{D}_2)$ are \textit{value-equivalent} iff their distinguished attributes are equal, i.e., they both represent the same edge or the same path possibly with different validity intervals. Formally, $t_1 = t_2 \Leftrightarrow u_1 = u_2, v_1 = v_2, l_1= l_2$.
\end{definition}

\textit{Value-equivalence} is used for temporal coalescing of tuples with adjacent or overlapping validity intervals \cite{lingozsu09}.
We extend the \textit{coalesce} primitive from temporal database literature~\cite{clifford1994consensus} to sgts with an aggregation function over the non-distinguished payload attribute, $\mathcal{D}$, as shown below:

\begin{definition}[Coalesce Primitive]\label{def:coalesce}
\sloppy
The coalesce primitive merges a set of value-equivalent sgts $\{ t_1, \cdots, t_n\}$, $t_i = (src, trg, l, [ts_i, exp_i), \mathcal{D}_i)$ for $1 \leq i \leq n$ with overlapping or adjacent validity intervals using an operator-specific aggregation function $f_{agg}$ over the payload attribute $\mathcal{D}$:
\begin{align*}
    & \text{coalesce}_{f_{agg}}(\{t_1 \cdots, t_n\}) = \\ 
    & (src, trg, l, [\min_{1 \leq i \leq n}{(ts_i)}, \max_{1 \leq i \leq n}{(exp_i)}), f_{agg}(\mathcal{D}_1, \cdots, \mathcal{D}_n) )
\end{align*}
\end{definition}

Distinguished attributes $src, trg$ and the label $l$ of sgts in a streaming graph $S$ represent the topology of a materialized path graph.
Hence, a finite subset of a streaming graph $S$ corresponds to a materialized path graph over the set of edges and paths that are in the streaming graph and the set of vertices that are adjacent to these.
We now use this to define snapshot graphs and the property of \textit{snapshot reducibility}.

\begin{definition}[Snapshot Graph]\label{def:snapshotgraph} 
A snapshot graph $G_{t}$ of a streaming graph $S$ is defined by a mapping  from each time instant in $\mathcal{T}$ to a finite set of sgts in $S$.
At any given time $t$, the content of a mapping $\tau_t(S)$ defines a snapshot graph 
$G_{t} = (V_{t}, E_{t}, P_{t}, \Sigma_{t}, \psi, \rho, \phi)$ where $E_{t} = \{ e_i \mid e_i.ts \leq t < e_i.exp \}$ 
is the set of all edges that are valid at time $t$, $P_{t} = \{ p_i \mid p_i.ts \leq t < p_i.exp \}$, and $V_{t}$ is the set of all vertices that are endpoints of edges and paths in $E_{t}$ and $P_{t}$, respectively.
\end{definition}

Value-equivalence (Definition \ref{def:equivalence}) and the coalesce primitive (Definition \ref{def:coalesce}) ensure that snapshot graphs have the \textit{set semantics}, i.e., at any point in time $t$, the snapshot graph $G_{t}$ of a streaming graph $S$, a vertex, edge and path exists at most once.

\section{Streaming Graph Queries}
\label{sec:query-system}

This section presents our \textit{streaming graph query} (SGQ) model.
We first provide a formal definition of SGQ using Datalog, enabling the specification of precise SGQ semantics  and to reason about its expressiveness. We then describe how SGQ captures a significant subset of existing graph query languages and provide concrete examples on how to formulate SGQ using a slight extension of G-CORE.

\subsection{Query Model}
\label{sec:query-model}

We formally describe SGQ based on a streaming generalization of the Regular Query (RQ) model \cite{reutter2017regular}.
Informally, RQ corresponds to \textit{binary, non-recursive} subset of Datalog with transitive closure
and provides a principled way to combine subgraph patterns and path navigations.
RQ provides a good basis for building a general-purpose framework for persistent query evaluation over streaming graphs, because (i) unlike UCRPQ, it is closed under transitive closure and therefore composable, (ii) it has more expressive power than the existing graph query languages such as SPARQL v1.1, Cypher, PGQL -- RQ strictly subsumes UCRPQ on which these are based, and (iii) its query evaluation and containment complexity is reasonable \cite{reutter2017regular}.
Due to its well-defined semantics and computational behaviour, RQ has been gaining popularity as a logical foundation for graph queries, both in theory \cite{bonifati2018querying, bonifati2019graph} and in practice \cite{angles2018g}.
Indeed,
RQ captures the core of the contemporary graph query language G-CORE that we use throughout this paper.

\begin{definition}[Regular Queries (RQ) -- Following \cite{reutter2017regular}]\label{def:rq}
The class of Regular Queries is the subset of non-recursive Datalog with a finite set of rules where each rule is of the form:
\footnote{The \textit{dependency graph} of a Datalog program is a directed graph whose vertices are its predicates and edges represent dependencies between predicates, i.e., there is an edge from $p$ to $q$ if $q$ appears in the body of rule with head predicate $p$. A Datalog program is \textit{non-recursive} iff its dependency graph is acyclic, i.e., no predicate depends recursively on itself.}
\[
head \leftarrow body_1, \cdots, body_n
\]
Each $body_i$ is either (i) a binary predicate $l(src, trg)$ where  $l \in \Sigma$ is a label, or (ii) $(l^*(src, trg) \text{ as }d)$, which is a transitive closure over $l(src, trg)$ for a label $l\in \Sigma, d \in \Sigma \setminus \phi(E)$, and each head predicate ($head$) is a binary predicate with $d(src, trg)$ for a label $d \in \Sigma \setminus \phi(E)$  except the reserved predicate $Answer \not\in \Sigma$ of an arbitrary arity.
\end{definition}

In other words, an RQ is a binary, non-recursive Datalog program extended with the transitive closure of binary predicates where input graph edges with a label $l \in \phi(E^I)$ correspond to instances of the extensional schema (EDB) and derived edges and paths with a label $l \in \Sigma \setminus \phi(E^I)$ correspond to instances of the intensional schema (IDB).
EDBs are predicates that appear only on the right-hand-side of the rules, which correspond to stored relations in Datalog \cite{abiteboul1995foundations}.
Similarly, we define IDBs as predicates that appear in the rule heads, which correspond to output relations in Datalog.

\begin{example}[Regular Query]\label{ex:regular_query}
Consider the real-time notification query in Example \ref{ex:alert} and its graph pattern in Figure \ref{fig:running_example}.
The one-time query for the same graph pattern corresponds to the following RQ:
\begin{align*}
\small
    RL(u_1, u_2)  & \leftarrow l(u_1, m_1), f^+(u_1, u2)\ as\ FP, p(u_2, m_1) \\
    Notify(u, m) & \leftarrow RL^+(u, v)\ as\ RLP, p(v, m) \\
    Answer(u,m) & \leftarrow Notify(u,m)
\end{align*}
where predicates $l, f, FP, p, RL, RLP$ represent labels {\normalfont \pred{likes}}, {\normalfont \pred{follows}}, {\normalfont \pred{followsPath}}, {\normalfont \pred{post}}, {\normalfont \pred{recentLiker}} and {\normalfont \pred{recentLikerPath}}, respectively.
\end{example}

Next, we define the notion of snapshot reducibility that enables us to precisely define the semantics of streaming queries and operators using their non-streaming counterparts.
Snapshot reducibility is used in temporal databases to generalize non-temporal queries and  operators to temporal ones \cite{clifford1994consensus}.

\begin{definition}[Snapshot-Reducibility]\label{def:snapshot-reducibility}
Let $S$ be a streaming graph, $\mathcal{Q}$ a streaming graph query and $\mathcal{Q}^O$ its non-streaming, one-time counterpart.
Snapshot reducibility states that each snapshot of the result of applying $\mathcal{Q}$ to $S$ is equal to the result of applying its non-streaming version $\mathcal{Q}^O$ over the corresponding snapshots of $S$, i.e., $\forall t \in \mathcal{T}, \tau_t\big(\mathcal{Q}(S)\big) = \mathcal{Q}^O\big(\tau_t(S)\big) $.
\end{definition}

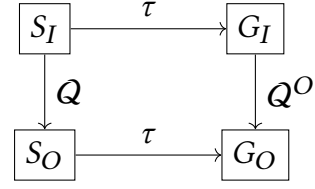
\begin{figure}
    \centering
    \begin{tikzpicture}[scale=1.4, transform shape]
            \node[draw] (si) {$S_I$};
            \node[draw] (so) at ($(si) - (0,1.2)$) {$S_O$};

            \node[draw] (gi) at ($(si) + (2,0)$) {$G_I$};
            \node[draw] (go) at ($(gi) - (0,1.2)$) {$G_O$};

            \path[->] (si) edge node[] [right] {$\mathcal{Q}$ } (so);
            \path[->] (gi) edge node[] [right] {$\mathcal{Q}^O$ } (go);
            \path[->] (si) edge node[] [above] {$\tau$} (gi);
            \path[->] (so) edge node[] [above] {$\tau$} (go);
        \end{tikzpicture}
    \caption{Snapshot reducibility (adapted from \cite{kramer2009semantics}).}
    \label{fig:snapshot-reducibility}
\end{figure}

Following existing research \cite{golabo03a}, we define the semantics of persistent evaluation of SGQ using the notion of snapshot reducibility (Definition \ref{def:snapshot-reducibility}).
It is known that for many operations such as joins and aggregation, exact results cannot be computed with a finite memory over unbounded streams \cite{bbd+02}.
In streaming systems, a common solution for bounding the space requirement is to evaluate queries on a window of data from the stream.
The windowed evaluation model provides a tool to process unbounded streams with bounded memory, and restricts the scope of queries to recent data, a desired feature in many applications\cite{golabo03a, bbd+02}.
Additionally, as opposed to streaming approximation techniques that trade off exact answers in favour of bounding the space requirements, window-based query evaluation enables exact query answers w.r.t. window specifications.
Hence, we adopt the \textit{time-based sliding window} for SGQ in the remainder.

\begin{definition}[Streaming Graph Query -- SGQ]
\label{def:streaming-query}
An SGQ query $Q$ is an RQ defined over a streaming graph $S$ and a time-based sliding window $\mathcal{W}_{\mathcal{T}}$ whose semantics is defined using the corresponding, one-time RQ $Q^O$ and the notion of snapshot reducibility (Definition \ref{def:snapshot-reducibility}): 
\begin{align*}
 \forall t \in \mathcal{T}, \quad \tau_t\big(  Q(S,\mathcal{W}_{\mathcal{T}}) \big) = 
 Q^O\Big( \tau_t \big( \mathcal{W}_{\mathcal{T}}(S) \big) \Big)
\end{align*}
\end{definition}

Figure \ref{fig:snapshot-reducibility} illustrates the correspondence between streaming and one-time graph queries. 
A direct consequence of such a relationship is that SGQ can be evaluated by repeatedly executing the corresponding one-time query, known as \textit{query re-evaluation} \cite{ghanem2006incremental}.
Specifically, the resulting streaming graph of an SGQ
can be obtained from the sequence of snapshots that is the result of repeated evaluation of the corresponding one-time query at every time instant:
an sgt $(u, v, l, [ts, expiry), D)$ is in the resulting streaming graph for an SGQ $Q$ $\tau_t \big( Q(S,\mathcal{W}_{\mathcal{T}})\big)$ if there is an edge $e=(u,v,l)$ or a path $p: u \stackrel{p}{\rightarrow} v$ with $l = \phi^p(p)$ in the resulting snapshot graph of the corresponding one-time query $G_{t_i} = Q^O\Big( \tau_{t} \big( \mathcal{W}_{\mathcal{T}}(S) \big)  \Big)$ for $ts \leq t < exp$.
However, such a strategy is wasteful as the input differences between two consecutive instants are likely to be small.
Alternatively, \textit{incremental evaluation} computes the changes in the output as new sgts arrive and old sgts expire due to window movements.
In this paper, we focus on the \textit{incremental evaluation} method and use the concept of \textit{snapshot reducibility} to ensure correct evaluation semantics.

\subsection{SGQ in Practice}
\label{sec:gcore}

The SGQ model formalizes an important class of graph queries in the streaming model.
It captures the core features of current graph query languages such as subgraph pattern and reachability-based path queries.
In this section, we demonstrate the SGQ's expressive power by mapping core G-CORE constructs to SGQ.
We choose G-CORE for demonstration due to the following reasons. G-CORE fulfills all three requirements of graph querying (\textbf{R1}, \textbf{R2} \& \textbf{R3} in Section \ref{sec:intro}). Other existing languages (e.g., SPARQL v1.1, Cypher, PGQL) can only partially satisfy these requirements due to (i) the lack of algebraic closure and composability, and (ii) limited path navigation capability \cite{bonifati2018querying}. Moreover, G-CORE supports SGQ capabilities such as the treatment of paths as first-class citizens and returning graphs. Finally, G-CORE is one of the more prominent  language specifications influencing the ongoing standardization process of a graph query language GQL (\url{https://www.gqlstandards.org}). 
\color{black}

Since we have not yet included properties in the data model, we focus on a subset of G-CORE where queries do not contain predicates or aggregation over property values.
This particular subset already covers many important key features: \textbf{(a)} returning graphs, \textbf{(b)} ASCII-art syntax for pattern matching, \textbf{(c)} joins over multiple graphs, \textbf{(d)} views and optionals, \textbf{(e)} RPQ-based reachability queries, \textbf{(f)} and powerful path patterns as demonstrated in the two examples we discuss below. 

G-CORE is originally targeted for one-time queries over static property graphs and it does not provide native windowing constructs.
We slightly extend the \lstinline{ON} clause with a \lstinline{WINDOW} clause to incorporate window specifications.
In particular, 
a time-based sliding window is defined by the newly introduced \lstinline{WINDOW} clause that specifies the window length, and an optional \lstinline{SLIDE} clause that specifies the slide interval, following a streaming graph reference in the \lstinline{ON} clause. 


\begin{figure}
    \centering
    \begin{lstlisting}%[
%    language=SQL,
%    showspaces=false,
%    emphstyle={\color{blue}\bfseries},
%    emph={PATH, CONSTRUCT, MATCH, WHERE, WINDOW, ON},
%    basicstyle=\ttfamily\small,
%    numbers=none,
%    commentstyle=\color{gray}]

PATH RL = (u1)-/ <:follows^*> /->(u2),
          (u1)-[:likes]->(m1)<-[:posts]-(u2)
CONSTRUCT (u) -[:notify]-> (m)
MATCH (u)-/ p<~RL*> /->(v),
      (v)-[:posts]->(m),
ON social_stream WINDOW(24h) SLIDE(1h)
\end{lstlisting}
    \caption{G-CORE representation of the SGQ in Example \ref{ex:alert}.}
    \label{fig:gcore}
\end{figure}

\begin{figure}
    \centering
    \begin{lstlisting}%[
%    language=SQL,
%    showspaces=false,
%    emphstyle={\color{blue}\bfseries},
%    emph={PATH, CONSTRUCT, MATCH, WHERE, WINDOW, ON, GRAPH, VIEW, AS, OPTIONAL},
%    basicstyle=\ttfamily\small,
%    numbers=none,
%    commentstyle=\color{gray}]
    
GRAPH VIEW rec_stream AS (
  CONSTRUCT (u1) -[:recommendation]-> (p)
  MATCH (u1)
    OPTIONAL (u1)-[:follows]->(u2)
    OPTIONAL (u1)-[:likes]->(m)<-[:posts]-(u2)
  ON social_stream WINDOW(24 hours) 
  MATCH (c) -[:purchase]->(p)
  ON tx_stream WINDOW(30d) SLIDE(1d)
  WHERE (u2) = (c) )
\end{lstlisting}
    \caption{G-CORE representation of the query in Example \ref{ex:rec-gcore}}
    \label{fig:gcore_integration}
\end{figure}

\begin{example}\label{ex:alert-gcore}
The G-CORE query in Figure \ref{fig:gcore} represents the real-time notification example in Example \ref{ex:alert} (its corresponding RQ is already given in Example \ref{ex:regular_query}).
Its {\normalfont \lstinline{PATH}} and {\normalfont \lstinline{MATCH}} clauses use ASCII-art syntax (\textbf{b}) to define complex graph patterns (\textbf{f}) with RPQ-based reachability (\textbf{e}), and its {\normalfont \lstinline{CONSTRUCT}} clause returns a streaming graph of {\normalfont \textsf{\small notify}} edges (\textbf{a}).
\end{example}

\begin{example}\label{ex:rec-gcore}
Consider the G-CORE query in Figure \ref{fig:gcore_integration} that combines streaming information from a social network of user interactions and a transaction network of customer purchases to drive product recommendations.
Its defines a view over the resulting streaming graph of  {\normalfont \pred{recommendation}} edges (\textbf{d})  by joining patterns from two streaming graphs (\textbf{c}), and its {\normalfont \lstinline{MATCH}} clause features optional predicates to incorporate two alternative social interactions (\textbf{d}).
Its graph pattern corresponds to the following RQ:
\begin{align*}
\small
    ACQ(u_1, u_2)  & \leftarrow l(u_1, m_1), p(u_2, m_1) \\
    ACQ(u_1, u_2) & \leftarrow f(u_1, u2) \\
    REC(u, p) & \leftarrow ACQ(u_1, u_2), pur(u_2, p) \\
    Answer(u,p) & \leftarrow REC(u,p)
\end{align*}
where predicates $l, f, p, pur, ACQ, REC$ represent labels \pred{likes}, \pred{follows}, \pred{post}, \pred{purchase}, \pred{acquaintance}, and \pred{recommendation}, respectively.
\end{example}


\section{Streaming Graph Algebra}
\label{sec:logical-plans}
This section presents the logical foundation of our streaming graph query processing framework.
We first introduce streaming graph algebra (SGA) and the semantics of its operators (Section \ref{sec:algebra}).
The main motivation behind the proposed SGA is similar to that of relational DBMSs: it enables us to formulate and represent query plans independent of specific physical implementations.
We subsequently describe how to transform a given SGQ (Definition \ref{def:streaming-query}) into its canonical SGA expression and illustrate logical query plans (Section \ref{sec:query-formulation}) and prove its closedness and composability (Section \ref{sec:algebra-composability}).
Finally, we present some transformation rules for SGA to demonstrate alternative logical plan generation opportunities (Section \ref{sec:query-transformation}).

\subsection{SGA Operators}
\label{sec:algebra}


For ease of exposition, the rest of the paper assumes that inputs to each SGA operator are partitioned into one more streaming graphs $S_a$ based on tuple labels where each $S_{a}$ contains sgts with the same label $a \in \Sigma$.
The output of each operator is also a streaming graph $S_{o}$ where each sgt has the label $o \in \Sigma \setminus \phi(E^I)$.\footnote{$\phi(E^I) \subset \Sigma$ is reserved for input graph edges and cannot be used by operators as labels for resulting sgts. In other words, $\phi(E^I) \subset \Sigma$ corresponds to EDBs in Datalog as described in Section \ref{sec:query-system}.}
SGA contains the following operators: windowing (Definition \ref{def:windowing}), filter (Definition \ref{def:selection}), union (Definition \ref{def:union}), subgraph pattern (Definition \ref{def:subgraph-pattern-operator}), and path navigation (Definition \ref{def:path-navigation-operator}).

\begin{definition}[\pred{WSCAN}]\label{def:windowing}
The \textbf{windowing  operator} $\mathcal{W}$ transforms a given input graph stream $S^I$ to a streaming graph $S$ by adjusting the validity interval of each sgt based on the window size $\mathcal{T}$ and the optional slide interval $\beta$, i.e., 
$\mathcal{W}_{\mathcal{T}, \beta}(S^I) := S: \big[ (u, v, l, [t, exp), \mathcal{D}: e(u, v, l)) \mid (u, v, l, t) \in S^I \wedge exp= \lfloor t / \beta \rfloor \cdot \beta + \mathcal{T} \big]$.
\end{definition}

The window size $\mathcal{T}$ determines the length of the validity interval of sgts and the slide interval $\beta$ controls the granularity at which the time-based sliding window progresses \cite{abw02, Pacaci_2020}. 
If $\beta$ is not provided, default is $\beta=1$, i.e., single time instant with the smallest granularity, and it defines a sliding window that progresses at every time instant.

The \pred{WSCAN} operator defines the semantics of time-based sliding windows. 
It acts as an interface between the external streaming graph sources and the query plans and it 
is responsible for providing data from input graph streams to a query plan, similar to the \textit{scan} operator in relational systems.
\pred{WSCAN} manipulates the implicit temporal attribute and associates a time interval to each sgt representing its validity.
Our model of representing  streaming graphs (Definition \ref{def:streaming-graph}) provides a concise representation of validity intervals and enables operators to treat time differently than the data stored in the graph.
This enables us to distinguish operator semantics from window semantics and eliminates the redundancy caused by integrating sliding window constructs into each operator of the algebra.
SGA operators access and manipulate validity intervals implicitly, generalizing their non-streaming counterparts with implicit handling of time.

\begin{example}\label{ex:windowing-operator}
Consider the real-time notification task of Example \ref{ex:alert} with a 24-hour window of interest.
{\normalfont \pred{WSCAN}} $\mathcal{W}_{24}$ sets validity intervals of sges of the input graph stream and produces a streaming graph where each sgt is valid for 24 hours, as shown in Figure \ref{fig:streaming-graph}.
\end{example}

\begin{definition}[\pred{FILTER}]\label{def:selection} \textbf{Filter operator}
$\sigma_{\Phi}(S)$  is defined over a streaming graph $S$ and a boolean predicate $\Phi$ involving the distinguished attributes of sgts, and its output stream consists of sgts of $S$ on which $\Phi$ evaluates to true.
Formally:
\begin{align*}
\small
\sigma_{\Phi}(S) = \big[ (u, v, l, [ts, exp), \mathcal{D}) \mid &
\\  (src, trg, l, [ts, exp), \mathcal{D}) \in S  &\land \Phi((src, trg, l, \mathcal{D}))
\big].
\end{align*}
\end{definition}


\begin{definition}[\pred{UNION}]\label{def:union} Union
$\cup^{[d]}$ with an optional output label $d \in \Sigma \setminus \phi(E^I)$ merges sgts of two streaming graphs $S_1$ and $S_2$, and assigns the new label $d$ if provided.
Formally:
\begin{align*}
    S_1 \cup^{[d]} S_2 & = \big[ t \mid t \in S_1 \lor t \in S_2    \big]
\end{align*}
\end{definition}


\begin{definition}[\pred{PATTERN}]\label{def:subgraph-pattern-operator}
The streaming \textbf{subgraph pattern operator} is defined as $\Join_{\Phi}^{src, trg, d}(S_{l_1}, \cdots, S_{l_n})$ where each $S_{l_i}$ is a streaming graph,
$\Phi$ is a conjunction of a finite number of terms in the form $pos_i = pos_j$ for $pos_i, pos_j \in \{ src_1, trg_1, \cdots, src_n, trg_n \}$ where $src_i, trg_i$ are endpoints of sgts in $S_{l_i}$, and 
$src, trg \in \{ src_1, trg_1, \cdots, src_n, trg_n \}$ are the endpoints of resulting sgts, 
and $d \in \Sigma \setminus \phi(E^I)$ represent the label of the resulting sgts.
Formally:
\begin{align*}
\Join_{\Phi}^{src, trg, d}&(S_{l_1}, \cdots, S_{l_n}) =  \big[ (u, v, d, [ts, exp), \mathcal{D}:e(u,  v, l)) \mid
\\ & \exists t_i = (src_i, trg_i, l_i, [ts_i, exp_i), \mathcal{D}_i) \in S_{l_i}, 1 \leq i \leq n 
\\ & \land \Phi((src_1, trg_1, \cdots, src_n, trg_n)) \land 
\\ & u = src \land v = trg \land \bigcap_{1 \leq i \leq n}{[ts_i, exp_i)} \neq \emptyset \land
\\ & ts = \max_{1 \leq i \leq n}{(ts_i)} \land exp = \min_{1 \leq i \leq n}{(exp_i)}
\big].
\end{align*}

\end{definition}

Given a subgraph pattern expressed as a conjunctive query, \pred{PATTERN} finds a mapping from vertices in the stream to free variables where (i) all query predicates hold over the mapping, and (ii) there exists a time instant at which each edge in the mapping is valid.

\begin{example}\label{ex:subgraph-operator}
Consider the real-time notification query given in Example \ref{ex:alert}; the {\normalfont \pred{recentLiker}} relationship defined in the form of a triangle  pattern can be represented with {\normalfont \pred{PATTERN}} $\Join_{\phi}^{src1, src4, RL}$ where $\phi = (trg_1 = trg_2 \land src_1 = src_3 \land src_2 = trg_3 )$.
Its output over the streaming graph, given in Figure \ref{fig:streaming-graph}, consists of sgts $ (y,RL,u, [28, 37), (y,RL,u)) $ and $ (u,RL,v, [29, 31), (u,RL,v))$ that correspond to derived edges with label {\normalfont \pred{recentLiker}}.
\end{example}

SGA operators may produce multiple value-equivalent sgts with adjacent or overlapping validity intervals. 
Unless otherwise specified, such sgts in resulting streaming graphs of SGA operators are coalesced to maintain the set semantics of streaming graphs and their snapshots (Definition \ref{def:equivalence}).
To illustrate, consider \pred{PATTERN} in the above example: 
over the streaming graph given in Figure \ref{fig:streaming-graph}, the \pred{PATTERN} operator finds two distinct subgraphs with vertices $(u,b,v)$ and $(u,c,v)$.
Consequently, it produces two value-equivalent tuples $ (u,RL,v, [29, 31), (u,RL,v))$ and $ (u,RL,v, [30, 31), (u,RL,v))$, which are coalesced into a single sgt by merging their validity intervals.

\begin{definition}[\pred{PATH}] \label{def:path-navigation-operator}
The streaming \textbf{path navigation operator} is defined as $ \mathcal{P}_R^d  (S_{l_1}, \cdots, S_{l_n})$ where $R$ is a regular expression over the alphabet $\{ l_1, \cdots, l_n \} \subseteq \Sigma$, and $d \in \Sigma \setminus \phi(E^I)$ designates the label of the resulting sgts.
The sgt $t=(u,v, l, [ts, exp), \mathcal{D}:p)$ is an answer for $\mathcal{P}_R^l $ if there exists a path $p$ between $u$ and $v$ in the snapshot of $S$ at time $t$, i.e., $p: u \stackrel{p}{\rightarrow} v \in \tau_{t}(S) = G_{t}$, and the label sequence of the path $p$, $\phi^p(p)$ is a word in the regular language $L(R)$.
Formally: 
\begin{align*}
    \mathcal{P}_R^d  & (S_{l_1}, \cdots, S_{l_n}) = \big[ (u, v, d, [ts, exp), \mathcal{D}) \mid   \exists p: u \stackrel{p}{\rightarrow} v \land
    \\ & \forall e_i \in p, \exists t_i = (src_i, trg_i, l_i, [ts_i, exp_i), \mathcal{D}_i) \in S_{l_i} \land
    \\ &   \phi^p(p) \in L(R) \land \bigcap_{t \in p}{[t.ts, t.exp)} \neq \emptyset  \land
    \\ &  ts = \max_{t \in p}{(t.ts)} \land  exp = \min_{t \in p}{(t.exp)} \land \mathcal{D} = p \big].
\end{align*}
\end{definition}

\pred{PATH} finds pairs of vertices that are connected by a path where (i) each edge in the path is valid at the same time instant, and (ii) path label is a word in the regular language defined by the query. 
This closely follows the RPQ model where path constraints are expressed using a regular expression over the set of labels \cite{wood2012query}.
Path navigation queries in the RPQ model are evaluated
under \textit{arbitrary} and \textit{simple} path semantics.
The former allows a path to traverse the same vertex multiple times, whereas under the latter semantics a path cannot traverse the same vertex more than once  \cite{angles2017foundations, wood2012query, baeza13}.
In this paper, we adopt the arbitrary path semantics due to its widespread adoption in modern graph query languages \cite{angles2018g, angles2017foundations, grades2016_van2016pgql}, and the tractability of the corresponding evaluation problem \cite{baeza13}.

\begin{example}\label{fig:path-operator}
In the example of Figure \ref{ex:alert} the path navigation over the derived {\normalfont \pred{recentLiker}} edges is represented by {\normalfont \pred{PATH}} $\mathcal{P}_{RL^+}^{RLP}$.
Its output over the resulting streaming graph of {\normalfont \pred{PATTERN}} of Example \ref{ex:subgraph-operator} consists of sgts $(y,RLP,u, [28, 37), (y,RL,u))$, $(u,RLP,v, [29, 31), (u,RL,v))$, and $(y,RLP,v, [29, 31),$ $\langle (y,RL,u), (u, RL, v) \rangle)$ that correspond to materialized paths with label {\normalfont \pred{recentLikerPath}} of length one and two.
\end{example}

Most existing work on the RPQ model focuses on the problem of determining reachability between pairs of vertices connected by a path conforming to given regular expression \cite{mendelzon1995finding, libkin2012regular, koschmieder2012regular, Pacaci_2020}.
By adapting the materialized path graph model (Definition \ref{def:path-graph}), we pinpoint that \pred{PATH} is equipped with the ability to return paths, i.e., each resulting sgt contains the actual sequence of edges that form the path with a label sequence conforming to given regular expression.

SGA builds on the Regular Property Graph Algebra (RPGA) \cite{bonifati2018querying}, which is itself based on Regular Queries (RQ). Of course, both RPGA and RQ formulate graph queries over static graphs, while SGA  operators are defined over streaming graphs (Definition \ref{def:streaming-graph}), and they access and manipulate validity intervals implicitly. Thus they generalize their non-streaming counterparts with implicit handling of time.
This follows from the fact that the semantics of SGA operators are defined through \textit{snapshot reducibility} (Definition \ref{def:snapshot-reducibility}), that is, the snapshot of the result of a streaming operator on a streaming graph $S$ at time $t$ is equal to the result of the corresponding non-streaming operator on the snapshot of the streaming graph $S$ at time $t$.


\subsection{Formulating Query Plans in SGA}
\label{sec:query-formulation}

SGA can express all queries that can be specified by SGQ (Section \ref{sec:query-system}).
This section provides an algorithm for the conversion. 

Given a SGQ $Q(S, \mathcal{W}_{\mathcal{T}})$ over a streaming graph and a time-based sliding window definition, 
Algorithm \textbf{\ref{alg:sga-generator}} produces
the canonical SGA expression.
The algorithm processes the predicates of a given SGQ and generates the corresponding SGA expression in a bottom-up manner:
each EDB $l$ corresponds to a \pred{WSCAN} over an input streaming graph $S^I_l$, each application of transitive closure corresponds to a \pred{PATH}, each IDB $d$ corresponds to a \pred{UNION} or \pred{PATTERN} based on the body of the corresponding rule.


\begin{algorithm}
\small
\SetAlgoRefName{SGQParser}
	\SetKwData{Left}{left}\SetKwData{This}{this}\SetKwData{Up}{up}
	\SetKwInOut{Input}{input}\SetKwInOut{Output}{output}
	\Input{ 
	Streaming Graph Query $Q(S, \mathcal{W}^{\mathcal{T}})$
	}
	\Output{SGA Expression $e$}

    $G_{Q} \leftarrow \texttt{Graph}(Q)$ \tcp{dependency graph } \label{line:sga-dependency}
    $[r_1, \cdots, r_n] \leftarrow \texttt{TopSort}(G_Q)$ \tcp{topological sort} \label{line:sga-topological}
    $exp \leftarrow []$ \tcp{empty mapping} \label{line:sga-init-mapping}
    
    \For{$1 \leq i \leq n$}{ \label{line:sga-visit-order}
    
        \Switch{$r_i$}{
            \Case{$l(src, trg), l \in \phi(E^I)$}{ $exp[l] \leftarrow \mathcal{W}^{\mathcal{T}}(S_l)$} \label{line:sga-edb}
            
            \Case{$l^*(x,y) as d$}{$exp[d] \leftarrow \mathcal{P}^d_{l*}(exp[l])$} \label{line:sga-path}
            
            \Other{\label{line:sga-other}
                $d(src, trg) \leftarrow r_i.head$, 
                $[b_1, \cdots, b_n] \leftarrow r_i.body$ \\
                $\Phi \leftarrow \texttt{GenPred}(r_i.body)$ \label{line:sga-generate-condition} \\
                $e \leftarrow \Join_{\Phi}^{src, trg, d}(exp[b_1], \cdots, exp[b_n] )$ \label{line:sga-generate-join}\\ 
                \If{$exp[d] \neq \emptyset$}{\label{line:sga-union}
                    $exp[d] \leftarrow exp[d] \cup e$ 
                }
                
                \Else{$exp[d] \leftarrow  e$ \label{line:sga-conjunction} }
            }
        }
    }
    
    \Return{$exp[Answer]$\label{line:sga-return}}

	\caption{}\label{alg:sga-generator}
\end{algorithm} 

    
    
    

\begin{theorem}\label{thm:expressiveness}
There exists a SGA expression $e \in SGA$ for any $Q \in SGQ$.
\end{theorem}

\begin{proof}
The dependency graph of an RQ is acyclic as RQ is non-recursive (Definition \ref{def:rq}); hence, Line \ref{line:sga-topological} is guaranteed to define a partial order over $Q$'s predicates.
Algorithm \textbf{\ref{alg:sga-generator}} generates an SGA expression for each predicate in this order and caches it in $exp$ array. 
In particular, Line \ref{line:sga-edb} generates an SGA expression for each EDB predicate and Line \ref{line:sga-path} generates a \pred{PATH} expression for each body predicate with a Kleene star.
For each rule $d(src, trg) := l_1(src_1, trg_1), \cdots, l_n(src_n, trg_n)$, Line \ref{line:sga-generate-join} generates a \pred{PATTERN} expression.
Finally, Line \ref{line:sga-union} generates a \pred{UNION} expression if there are multiple rules with the same head predicate $d$.
As each predicate is processed based on the partial order defined by the dependency graph $G_Q$ (Line \ref{line:sga-visit-order}), $exp$ is guaranteed to have SGA expressions for each predicate $r_j (1 \leq j \leq i)$ when processing predicate $r_i$.
Once all predicates are processed, Line \ref{line:sga-return} returns the SGA expression of the $Answer$ predicate.
Hence, Algorithm \textbf{\ref{alg:sga-generator}} correctly constructs an SGA expression for a given SGQ.
\end{proof}

The complexity of evaluating SGA expressions is the same as 
RQ given their relationship noted above: NP-complete in combined complexity and NLogspace-complete in data complexity \cite{reutter2017regular, bonifati2018querying}.

\begin{example}[Canonical Translation]\label{ex:canonical_form}
For the real-time notification task in Example \ref{ex:alert} and its corresponding RQ in Example \ref{ex:regular_query},
Algorithm \textbf{\ref{alg:sga-generator}} generates the following canonical SGA expression for its corresponding SGQ with a sliding window of 24 hours:
\begin{align*}
&     \Join_{\phi_2}^{(src1, trg2, notify)} \bigg( \\
        & \qquad \mathcal{P}_{RL^+}^{RLP} \Big( \Join_{\phi_1}^{src1, src2, RL} \big(\mathcal{W}_{24}(S_{l}), \mathcal{W}_{24}(S_{p}), 
        \mathcal{P}_{f^+}^{FP} \big (\mathcal{W}_{24}(S_{f}) \big) \big) \Big), \\
        & \qquad  \mathcal{W}_{24}(S_{p}) 
 \quad  \bigg) \\ 
& \phi_1 = (trg_1 = trg_2 \land src_1 = src_3 \land src_2 = trg_3 ), \\ 
& \phi_2 = (trg_1 = src_2) 
\end{align*}
\end{example}

\begin{figure}
\centering
    \includegraphics[width=\linewidth]{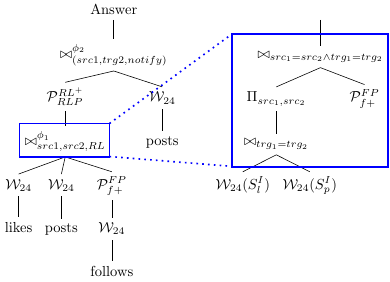}
    \caption{(left) Logical plan for the SGA expression in Example \ref{ex:canonical_form}, and (right) binary join tree for its \pred{PATTERN}.}
    \label{fig:sga-q6-plans}
\end{figure}

Figure \ref{fig:sga-q6-plans} (left) illustrates the logical plan for the same SGQ that consists of SGA operators.

\subsection{Closedness and Composability}
 \label{sec:algebra-composability}

Algebraic closure is a required property of any query algebra as it enables query rewriting (Section \ref{sec:query-transformation}) and query optimization. Composability is a desired feature for a declarative query language as it facilitates query decomposition, view-based query evaluation, query rewriting etc.
SGA operators are closed over streaming graphs as defined in Section \ref{sec:data-model}; that is, the output of an SGA operator is a valid streaming graph if its inputs are valid streaming graphs. 
Thus SGA queries are composable, i.e., the output of one query can be used as input of another query.

SGQ language is also closed (Theorem \ref{thm:expressiveness}) -- each query takes one or more streaming graphs as input and produces a streaming graph as output. 
It is also composable as the output of a query can be the input of the subsequent query.
As such, G-CORE variation that we use as our user-level query language example (Section \ref{sec:gcore}) attains composability exactly as its original version is composable over property graphs \cite{angles2018g}.
This is in contrast to the other graph query languages that lack an algebraic basis like ours, e.g., SPARQL and Cypher are not composable and may not be closed. 
Cypher 9 requires graphs as input, but produces tables as output so the language is neither closed nor composable -- the results of a Cypher query cannot be used as input to a subsequent one without additional processing. 
SPARQL can produce graphs as output using the \lstinline{CONSTRUCT} clause, and is therefore closed; however,  it requires query results to be made persistent and therefore not easily composable  \cite{bonifati2018querying}. 


\subsection{Transformation Rules}
\label{sec:query-transformation}

As noted above, closedness of an algebra is important for query rewriting to explore the space of equivalent plans; this is a key component of query planning and  optimization.
Although optimization problems are beyond the scope of this paper, in this section we highlight possible transformation rules that enable the systematic exploration of the plan space in order to demonstrate the potential of our SGA.

Some of the traditional relational transformation strategies such as join ordering, predicate push down are applicable to  \pred{UNION},  \pred{FILTER} and \pred{PATTERN} due to  snapshot-reducibility. \pred{UNION} and \pred{FILTER} operators are streaming generalizations of corresponding relational union and selection operators, and the \pred{PATTERN} operator can be represented using a series of equijoins.
Below, we describe transformation rules involving the other novel SGA operators:

\textbf{Transformation Rules for \pred{WSCAN}:}  \pred{WSCAN}  $(\mathcal{W}^{\mathcal{T}})$ commutes with operators that do not alter the validity intervals of sgts, i.e., \pred{UNION} and \pred{FILTER}.
Pushing \pred{FILTER} down the \pred{WSCAN} operator can potentially reduce the rate of sgts and consequently the amount of state the windowing operator needs to maintain.
Formally:
\begin{enumerate}
    \item $\mathcal{W}^{\mathcal{T}}(\sigma^{\phi}(S)) = \sigma^{\phi}(\mathcal{W}^{\mathcal{T}}(S))$
    \item $\mathcal{W}^{\mathcal{T}}(S_1 \cup S_2) = \mathcal{W}^{\mathcal{T}}(S_1) \cup \mathcal{W}^{\mathcal{T}}(S_2)$ 
\end{enumerate}

\textbf{Transformation Rules for \pred{PATH}:} We identify two transformation rules for \pred{PATH}:
\begin{enumerate}
	\item Alternation: $\mathcal{P}_{a \mid b}^d(S_a, S_b) = \bigcup^d(S_a, S_b)$
	\item Concatenation: $\mathcal{P}_{a \cdot b}^d(S_a, S_b) = \Join_{trg_1 = src_2}^{src1, trg2, d}(S_a, S_b)$
\end{enumerate}

These transformation rules enable the exploration of a rich plan space for SGQ that are represented by SGA.
In particular, \pred{PATH} and its transformation rules enable the integration of existing approaches for RPQ evaluation with standard optimization techniques such as join ordering and pushing down selection in a principled manner. 
Traditionally, path query evaluation follows two approaches \cite{Pacaci_2020, koschmieder2012regular, salihoglu2019graph, fan2017incremental, baeza13, yakovets2016query}: graph traversals guided by finite automata, or relational algebra extended with transitive closure, i.e., $alpha$-RA.
Yakovets et al. introduce a hybrid approach (Waveguide) and model the cost factors that impact the efficiency of RPQ evaluation on static graphs \cite{yakovets2016query}.
SGA enables the representation of these approaches in a uniform manner, and the above transformation rules enable the exploration of a rich plan space that subsumes these existing plans.
An example application of these transformation rules and a micro-benchmark demonstrating the potential benefits of query optimizations through plan space exploration is given in Section \ref{sec:experiments-transformation}.

    

\section{Implementation}
\label{sec:implementation}

We  implemented a prototype streaming graph query processor (\url{https://dsg-uwaterloo.github.io/s-graffito/}) based on the algebraic framework we propose in this paper.
Our goal is to build a plausible conceptual framework for expressing and evaluating SGQ -- not to have a complete system. Hence, we focus on the construction of query execution plans and physical implementations of logical SGA operators. 



\subsection{Query Processor}
\label{sec:query-processor}

Conceptually, SGQ can be evaluated by repeatedly evaluating from scratch the corresponding one-time query at each point in time (Section \ref{sec:query-model}).
Albeit semantically correct, such a re-execution strategy is, of course, infeasible.
In streaming systems, the focus is on \textit{incremental} evaluation of persistent queries where the goal is to avoid re-computing the entire result by only computing the changes to the output in real-time as new input arrives.
Consequently, queries are executed in a
data-driven (push-based) manner as opposed demand-driven (pull-based) query processing employed in traditional relational DBMSs \cite{graefe93}.
As such, one can implement the framework proposed in this paper over existing streaming systems such as Apache Flink, Spark Streaming and Timely Dataflow.
For our prototype, we use Timely Dataflow (TD) as the underlying execution engine.

Applications in TD are expressed as a directed graph of operations where vertices correspond to user-defined computations and edges correspond to the flow of data between them.
In executing an SGQ, the query processor first creates a logical plan from the canonical SGA expression of the given query.
The physical execution plan in the form of a dataflow graph is constructed by: (i) creating source vertices for leaves of the logical query plan that receives input graph streams, (ii) replacing logical SGA operators with physical operator implementations, and (iii) creating a sink vertex for the root of the logical query plan that pushes results back to the user.
Section \ref{sec:physical-algebra} describes physical operator implementations in detail.

TD associates each input data with a logical timestamp that enables fine-grained synchronization and progress tracking.
We represent each input graph stream as an evolving collection where each item represents an sge (Definition \ref{def:inputstream}) and event timestamps assigned by the source are used as logical timestamps.
Upon the arrival of a new edge, TD propagates the corresponding sge through the physical execution plan and computes the new output at the given logical timestamp.

\subsection{Physical Operator Algebra}
\label{sec:physical-algebra}

\subsubsection{Overview}

Physical operator implementations for streaming systems have two requirements: they should be push-based and non-blocking, so they do not need the entire input to be available before producing the first result.
The standard dataflow implementations of stateless \textsf{\small FILTER} and \textsf{\small UNION} operators can be directly used in SGA, and \textsf{\small WSCAN} can be implemented via the standard \textit{map} operator that adjusts the validity intervals of sgts based on window specifications.
We focus on the stateful operators \textsf{\small PATTERN} and \textsf{\small PATH} that need to maintain an internal operator state that is accessed during query processing. 
This state is updated as new sgts enter the window and old sgts expire.
As discussed earlier, time-based sliding windows ensure that the portion of the input that may contribute to any future result is finite, making incremental, non-blocking computation possible.

TD's Differential Dataflow (DD) layer ~\cite{cidr13_mcsherrymii13} provides a set of built-in, high-level programming primitives (operators) that can be used to compose arbitrary dataflows for general-purpose computations, and it automatically incrementalizes these.
Consequently, DD can be asked to evaluate SGQ by (i) creating a dataflow of DD operators for a given SGQ and (ii) maintaining the window content as an evolving collection.
Indeed, we use such a strategy as a competitive baseline in our experimental evaluation (Section \ref{sec:experiments_comparative}).
However, DD's generality comes at a performance cost for evaluating SGQ, as we show in Section \ref{sec:experiments_comparative}.
Below, we describe how to devise physical operator implementations specific to SGQ by utilizing the properties of the SGQ model.
Of course, these are not the only physical operator implementations that are possible for SGA; other implementations can possibly be developed, and these are exemplars to demonstrate the implementability of the SGA operators.
They are also what we use in the experiments.

\subsubsection{Pattern Matching}
\label{sec:pattern-operator}

Implementation of \textsf{PATTERN} models subgraph patterns as conjunctive queries that can be evaluated using a series of joins. There is a rich literature of streaming join implementations that can be used.
Symmetric hash join ~\cite{wilschut91}
is commonly used to implement non-blocking joins in the streaming model: a hash table is built for each input stream and upon arrival (expiration) of a tuple, it is inserted into (removed from) its corresponding hash table and other tables are probed for matches \cite{urhanf2000, vldb03_golabo03}.
This produces an append-only stream of results for internal windows that do not invalidate previously reported results upon expiration of their participating tuples \cite{golab:2006yq}.
For external windows that require eviction of old results as the windows slide forward, 
expired results can be determined by maintaining expiration timestamps: a join result expires when one of its participating tuples expire.
Our use of validity intervals enables the user or the application to adopt both window semantics without the need for explicit processing of expired input tuples.

Given a subgraph pattern, we take the standard approach of creating a binary join tree where leafs represent streaming graphs as input streams and internal nodes represent pipelined hash join operators.
For instance, Figure \ref{fig:sga-q6-plans} (right) shows the logical plan  for the query in Example \ref{ex:alert} and the join tree for its \textsf{\small PATTERN}.
In our prototype, we use the ordering of predicates in \textsf{\small PATTERN} to construct the join tree and leave the problem of finding efficient join plans (e.g. using  \textit{worst-case optimal joins} \cite{Ngo_2014}) for future investigation.

\subsubsection{Path Navigation}

DD's \textit{iterate} allows constructing cyclic dataflows that can model arbitrary nested iterations, 
and it can be used to evaluate \textsf{\small PATH} and its recursive path expressions.
However, the use of recursion in SGQ is limited to transitive closure (Section \ref{sec:query-system}), and the RPQ-based semantics of \textsf{\small PATH} is sufficient to evaluate this limited form of recursion (Theorem \ref{thm:expressiveness}).
In previous work~\cite{Pacaci_2020}, we study the design space of streaming algorithms for persistent RPQ  under arbitrary path semantics.
In brief, the streaming RPQ algorithm follows the automata-based RPQ evaluation method ~\cite{bonifati2018querying} and maintains a spanning forest-based data structure, called $\Delta$-tree index, that enables the compact representation of partial path segments.
In our prototype, we employ the streaming RPQ algorithm in \cite{Pacaci_2020} as the non-blocking, physical implementation of \textsf{\small PATH} operator, which eliminates the need for cycles in physical execution plans.
Additionally, a spanning-tree-based representation of intermediate results enables us to recover actual paths and allows queries to return and manipulate paths as first-class citizens.

\subsubsection{Customized \textsf{\small PATH} Implementation}
\label{sec:path-navigation-implementation}

We now describe a novel algorithm \textit{Streaming Path Navigation} (\textbf{\ref{alg:arbitrary}}) that can be used as an alternative physical operator for the \textsf{\small PATH} operator (Definition \ref{def:path-navigation-operator}). 
In contrast to our previous work \cite{Pacaci_2020}, \textbf{\ref{alg:arbitrary}} utilizes the validity intervals of path segments to simplify the state maintenance in the absence of explicit deletions.
Algorithms in \cite{Pacaci_2020} are based on the \textit{negative tuple} approach; expirations due to window movements are processed using the same machinery as explicit deletions.
Upon expiration (deletion) of an edge, their algorithm first finds all results that are affected by the expiration (deletion), then it traverses the snapshot graph to ensure that there is no alternative path leading to the same result.
This corresponds to re-derivation step of \textit{DRed} ~\cite{gms93}, optimized for RPQ evaluation on streaming graphs.
Instead, \textbf{\ref{alg:arbitrary}} utilizes the temporal pattern of sliding window movements and adopt the \textit{direct} approach, i.e., it can \textit{directly} determine expired tuples based on their validity intervals.
This is possible due to the separation of the implementation of sliding windows from operator semantics via an explicit \textsf{WSCAN} operator.


\begin{algorithm}
\small
\SetAlgoRefName{S-PATH}
	\SetKwData{Left}{left}\SetKwData{This}{this}\SetKwData{Up}{up}
	\SetKwInOut{Input}{input}\SetKwInOut{Output}{output}
	\Input{ 
	Input streaming graph $S$,
	Regular expression $R$,
	output label $o$
	}
	\Output{Output streaming graph $S_O$}
    
    $A(S, \Sigma, \delta, s_0, F) \leftarrow \texttt{ConstructDFA}(R)$ \\
    $\texttt{Initialize } \Delta-\texttt{PATH}$ \\
    $S_O \leftarrow \emptyset$ \\
    $\mathrm{R} \leftarrow \emptyset$ \\
    
    \ForEach{$(u,v,l, [ts, exp), \mathcal{D}) \in S$}{
        \ForEach{$s,t \in S$ where $t=\delta(s,l)$}{
            \If{$s = s_0 \wedge T_u \not\in \Delta-\texttt{PATH}$}{
                add $T_u$ with root node $(u,s_0)$
            }
            \If{$s=s_0$}{
                \If{$(v,t) \not\in T_u$}{
                    $\mathrm{R} \leftarrow \mathrm{R} +$ \textbf{\ref{alg:insertrapq_insert_edge}}($T_u$, $(u,s_0)$, $(v,t)$, $e(u,v)$) \label{line:insertrapq_callinsert_base}
                }
                \ElseIf{$(v,t).exp < exp$}{
                    $\mathrm{R} \leftarrow \mathrm{R} +$ \textbf{\ref{alg:updaterapq_update_subtree}}($T_u$, $(u,s_0)$, $(v,t)$, $e=(u,v)$) \label{line:updaterapq_callinsert_base}
                }
            }
            $\mathrm{T} \leftarrow $ \texttt{ExpandableTrees}($\Delta-\texttt{PATH}, (u,s), ts$) \label{line:expandable_trees} \\
            \ForEach{$T_x \in \mathrm{T} $}{
                \If{$(v,t) \not\in T_x$}{\label{line:insertrapq_check_existence}
                    $\mathrm{R} \leftarrow \mathrm{R} +$ \textbf{\ref{alg:insertrapq_insert_edge}}($T_x$, $(u,s)$, $(v,t)$, $e(u,v)$) \label{line:insertrapq_callinsert}
                }
                \ElseIf{$(v,t).exp < min((u,s).exp, exp)$}{\label{line:insertrapq_updateexpiry}
                    $\mathrm{R} \leftarrow \mathrm{R} +$ \textbf{\ref{alg:updaterapq_update_subtree}}($T_x$, $(u,s)$, $(v,t)$, $e=(u,v)$)
                }
            }
        }
    }
	\ForEach{sgt $t \in R$}{
	push $t$ to $S_O$
	}
    
	
	\caption{}\label{alg:arbitrary}
\end{algorithm} 

\begin{algorithm}
\small
\SetAlgoRefName{Expand}
	\SetKwData{Left}{left}\SetKwData{This}{this}\SetKwData{Up}{up}
	\SetKwInOut{Input}{input}\SetKwInOut{Output}{output}
	\Input{Spanning Tree $T_x$ rooted at $(x,s_0)$,parent $(u,s)$, \\ child $(v,t)$, edge $e(u,v)$ \\
	}
	\Output{Set of results $\mathrm{R}$}
	$\mathrm{R} \leftarrow \emptyset$ \\
    Insert $(v,t)$ as $(u,s)$'s child \\ \label{line:insertrapq_updateparent}
    $(v,t).ts = max(e.ts, (u,s).ts)$\\
    $(v,t).exp = min(e.exp, (u,s).exp)$ \\ \label{line:insertrapq_updatetimestamp}
    
    \If{$t \in F$}{
	    $p \leftarrow \texttt{PATH}(T_x, (v,t))$ \\
        $R \leftarrow R + (x,v,O, [(v,t).ts, (v,t).exp), p)$ \label{line:expand_addresult}
    }
    \ForEach{edge $e(v,w) \in G_{ts} $ s.t. $\delta(t, \phi(e))=q$}{\label{line:insertrapq_traverseedges}
        \If{$(w,q) \not\in T_x $}{ \label{line:insertrapq_exploreedge}
            $\mathrm{R} \leftarrow \mathrm{R} +$ \textbf{\ref{alg:insertrapq_insert_edge}}($T_x$, $(v,t)$, $(w,q)$, $e(v,w)$) \label{line:insertrapq_callinsert2}
        }
        \ElseIf{$(w,q).exp < min((v,t).exp, e.exp)$}{\label{line:expandrapq_updateexpiry}
        \label{line:insertrapq_updateedge}
            $\mathrm{R} \leftarrow \mathrm{R} +$ \textbf{\ref{alg:updaterapq_update_subtree}}($T_x$, $(v,t)$, $(w,q)$, $e(v,w)$)
        }
	}

    \Return $\mathrm{R}$
\caption{}\label{alg:insertrapq_insert_edge}
\end{algorithm}

\begin{algorithm}
\small
\SetAlgoRefName{Propagate}
	\SetKwData{Left}{left}\SetKwData{This}{this}\SetKwData{Up}{up}
	\SetKwInOut{Input}{input}\SetKwInOut{Output}{output}
	\Input{Spanning Tree $T_x$ rooted at $(x,s_0)$, parent $(u,s)$, \\
	child $(v,t)$, edge $e(u,v)$ \\
	}
	\Output{Set of results $\mathrm{R}$}
    $\mathrm{R} \leftarrow \emptyset$ \\
	$(v,t).pt = (u,s)$ \\
    $(v,t).ts = min((v,t).ts,  max(e.ts, (u,s).ts))$\\
	$(v,t).exp = max((v,t).exp, min(e.exp, (u,s).exp))$ \\
	\If{$t \in F$}{
	    $p \leftarrow \texttt{PATH}(T_x, (v,t))$ \\
        $\mathrm{R} \leftarrow \mathrm{R} + (x,v,O, [(v,t).ts, (v,t).exp), p)$ \label{line:propagate_addresult}
    }
	\ForEach{edge $e=(v,w) \in G_{ts}$ s.t. $\delta(t, \phi(e))=q $}{
	    \If{$(w,q).exp < min((v,t).exp, e.exp)$}{\label{line:propagate_updateexpiry}
	        $\mathrm{R} \leftarrow \mathrm{R} +$ \textbf{\ref{alg:updaterapq_update_subtree}}($T_x$, $(v,t)$, $(w,q)$, $e(v,w)$)
	    }
	}
	
	\Return $\mathrm{R}$
\caption{}\label{alg:updaterapq_update_subtree}
\end{algorithm}

Algorithm \textbf{\ref{alg:arbitrary}} incrementally performs a traversal of the underlying snapshot graph under the constraints of a given RPQ as sgts arrive.
It first constructs a DFA from the regular expression of a \textsf{PATH} operator, and initializes a spanning forest-based data structure, called $\Delta-\texttt{PATH}$, that is used as the internal operator state during query processing.
$\Delta-\texttt{PATH}$ is used to maintain a path segment, i.e., a partial result, between each pair of vertices in the form a spanning forest under the constraints of a given RPQ, consistent with Definition \ref{def:path-navigation-operator}.
Upon the arrival of an sgt, Algorithm \textbf{\ref{alg:arbitrary}} probes $\Delta-\texttt{PATH}$ to retrieve partial path segments that can be extended with the edge (or a path segment) of the incoming sgt.
Each partial path segment is extended with the incoming sgt, and Algorithm \textbf{\ref{alg:arbitrary}} traverses the snapshot graph $G_{t}$ until no further expansion is possible.

\begin{definition}[Spanning Tree $T_x$]\label{def:spanning-tree}
Given an automaton $A$ for the regular expression $R$ of a \textsf{PATH} operator $\mathcal{P}^R_d$ and a streaming graph $S$ at time $t$, a spanning tree $T_x$ forms a compact representation of valid path segments that are reachable from the vertex $x \in G_{t}$ under the constraints of a given RPQ, i.e., a vertex-state pair $(u,s)$ is in $T_x$ at time $t$ if there exists a path $p \in G_{t}$ from $x$ to $u$ with label $\phi^p(p)$ such that $s = \delta^*(s_0, \phi^p(p))$.
\end{definition}

A node $(u,s) \in T_x$ indicates that there is a path $p$ in the snapshot graph with label $\phi^p(p)$ such that $s = \delta^*(s_0, \phi^p(p))$, and this path can simply be constructed by following parent pointers ($(u,s).pt$) in $T_x$.
Under the arbitrary path semantics, there are potentially infinitely many path segments between a pair of vertices that conform to a given RPQ due to the presence of cycles in the snapshot graph and a Kleene star in the given RPQ.
Among those, \ref{alg:arbitrary} materializes the path segment with the largest expiry timestamp, that is, the path segment that will expire furthest in the future. 
Consequently, for each node $(u,s) \in T_x$, the sequence of vertices in the path from the root node to $(u,s)$ corresponds to the path from $x$ to $u$ in the snapshot graph with the largest expiry timestamp. 
This is achieved by the coalesce primitive (Definition \ref{def:coalesce}) with an aggregation function \texttt{max} over the expiry timestamp of path segments. \footnote{Arbitrary path semantics provides the flexibility for the aggregation function $f_{agg}$ of the coalesce primitive.}
Upon expiration of a node $(u,s)$ in $T_x$ and its corresponding path segment in the snapshot graph, this guarantees that there cannot be an alternative path segment between $x$ and $u$ that have not yet expired.
Hence, we can \textit{directly} find expired tuples based on their expiry timestamps.
This is based on the observation that expirations have a temporal order unlike explicit deletions, and \ref{alg:arbitrary} utilizes these temporal patterns to simplify window maintenance.

\begin{definition}[$\Delta-\texttt{PATH}$ Index]\label{def:delta}
Given an automaton $A$ for the regular expression $R$ of a \textsf{PATH} operator $\mathcal{P}^R_d$ and a streaming graph $S$ at time $t$, $\Delta-\texttt{PATH}$ is a collection of spanning trees (Definition \ref{def:spanning-tree}) where each tree $T_x$ is rooted at a vertex  $x \in G_{t}$ for which there is an sgt $t \in S(t)$ with a label $l$ such that $\delta(s_0, l) \neq \emptyset$ and $src=x$.
\end{definition}

$\Delta-\texttt{PATH}$ encodes a single entry for each pair of vertices under the constraints of a given query, consistent with the set semantics of snapshot graphs (Section \ref{sec:data-model}). 
Due to spanning-tree construction (Definition \ref{def:spanning-tree}), actual paths can easily be recovered by following the parent pointers; hence, $\Delta-\texttt{PATH}$ constitutes a compact representation of intermediate results for path navigation queries over materialized path graphs.
Our implementation models $\Delta-\texttt{PATH}$ as a hash-based inverted index from vertex-state pairs to spanning trees, enabling quick look-up to locate all spanning trees that contain a particular vertex-state pair.
Upon arrival of an sgt $t=(u,v,l, [ts, exp), \mathcal{D})$, Algorithm \textbf{\ref{alg:arbitrary}} probes this inverted index of $\Delta-\texttt{PATH}$ to retrieve all path segments that can be extended with the incoming sgt, that is, spanning trees that have the node $(u,s)$ with an expiry timestamp smaller than $ts$ for any state $s \in \{ s \in S \mid \delta(s,l) \neq \emptyset \}$ (Line \ref{line:expandable_trees}).
If the target node $(v,t)$ for $t=\delta(s,l)$ is not in the spanning tree $T_x$, Algorithm \textbf{\ref{alg:insertrapq_insert_edge}} is invoked to expand the existing path segment from $(x,0)$ to $(u,s)$ with the node $(v,t)$ and to create a new leaf node as a child of $(u,s)$.
In case there already exists a path segment between vertices $(x,0)$ and $(v,t)$ in $\Delta-\texttt{PATH}$, i.e., the target node $(v,t)$ is already in $T_x$, Algorithm \textbf{\ref{alg:arbitrary}} compares its expiry timestamp with the new candidate (Line \ref{line:insertrapq_updateexpiry}).
If the extension of the existing path segment from $(x,0)$ to $(u,s)$ with $(v,t)$ results in a larger expiry timestamp than $(v,t).exp$, Algorithm \textbf{\ref{alg:updaterapq_update_subtree}} is invoked to update the expiry timestamp of $(v,t)$ and its children in $T_x$.
Algorithms \textbf{\ref{alg:insertrapq_insert_edge}} and \textbf{\ref{alg:updaterapq_update_subtree}} traverse the snapshot graph until no further update is possible.
The following example illustrates the behaviour of Algorithm \textbf{\ref{alg:arbitrary}} on our running example.

\begin{figure*}
    \centering
    \begin{subfigure}[t]{0.35\textwidth}
        \includegraphics[width=\textwidth]{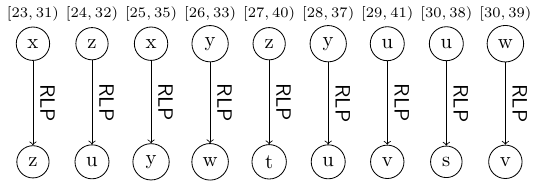}
        \caption{Streaming Graph $S$}
        \label{fig:stream-horizontal}
    \end{subfigure}
    \begin{subfigure}[t]{0.17\textwidth}
        \includegraphics[width=\textwidth]{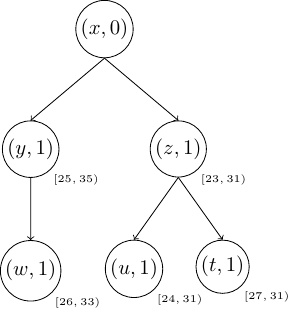}
        \caption{$t$ = 27}
        \label{fig:tree-v1}
    \end{subfigure}
        \begin{subfigure}[t]{0.18\textwidth}
        \includegraphics[width=\textwidth]{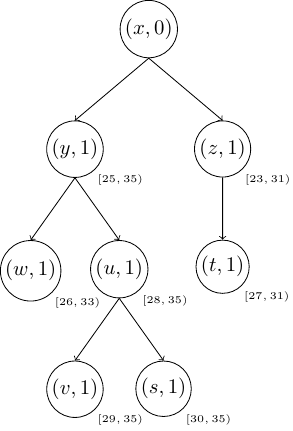}
        \caption{$t$ = 30}
        \label{fig:tree-v2-new}
    \end{subfigure}
    \begin{subfigure}[t]{0.18\textwidth}
        \includegraphics[width=\textwidth]{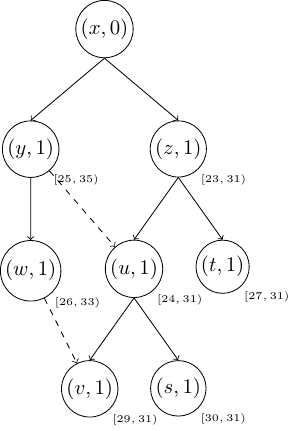}
        \caption{$t$ = 30 for \cite{Pacaci_2020}}
        \label{fig:tree-v2-old}
    \end{subfigure}
    \caption{(a) A streaming graph $S_{RLP}$ as the input for \textsf{PATH} operator, (b) spanning tree $T_x$ at $t = 28$, (c) spanning tree $T_x$ at $t = 30$ of the proposed algorithm following the \textit{direct} approach, and (d) spanning tree $T_x$ at $t = 30$ of \cite{Pacaci_2020} following the \textit{negative tuple} approach.}
    \label{fig:expiry-example}
\end{figure*}

\begin{example}\label{ex:navigation_algorithm}
Consider the real-time notification query in Example \ref{ex:alert} whose SGA expression is given in Example \ref{ex:canonical_form}.
Figure \ref{fig:stream-horizontal} shows an excerpt of the streaming graph input to the \textsf{PATH} operator $\mathcal{P}^{RL^+}_{RLP}$.
Figure \ref{fig:tree-v1} depicts a spanning tree $T_x \in \Delta-\texttt{PATH}$ at $t = 27$.
Upon arrival of the sgt $(y,u,RL,[28,37), \mathcal{D} = \{(y,RL,u)\})$ at $t=28$, Algorithm \textbf{\ref{alg:arbitrary}} extends the path segment from $(x,0)$ to $(y,1)$ with $(u,1)$, and compares its expiry timestamp with that of $(u,1)$ that is already in $T_x$. 
As the new extension has larger expiry timestamp, the validity interval and the parent pointer of $(u,1) \in T_x$ is updated (Line \ref{line:insertrapq_updateexpiry} in Algorithm \textbf{\ref{alg:arbitrary}}).
Then, incoming sgts at times $t=28$ and $t=29$ are processed by Algorithm \textbf{\ref{alg:insertrapq_insert_edge}} as corresponding target nodes $(v,1)$ and $(s,1)$ are not in $T_x$, adding $(v,1)$ and $(s,1)$ as children of $(u,1)$.
At $t=30$, the incoming sgt $(w,v,RL,[30,39), \mathcal{D} = \{(w,RL,v)\})$ might extend the path segment from $(x,0)$ to $(v,1)$ with expiry timestamp 33.
However, Algorithm \textbf{\ref{alg:arbitrary}} does not make any modification to $\Delta-\texttt{PATH}$, as $(v,1)$ is already in $T_x$ with a larger timestamp (Line \ref{line:insertrapq_updateexpiry}).
Figure \ref{fig:tree-v2-new} depicts the resulting spanning tree $T_x$ at $t=30$.
\end{example}

As described earlier, $\Delta-\texttt{PATH}$ stores a \textit{parent} pointer for each node pointing to its parent node in the corresponding spanning tree, and  
Algorithm \textbf{\ref{alg:updaterapq_update_subtree}} updates these pointers during processing.
By traversing these parent pointers for each resulting sgt, Algorithm \textbf{\ref{alg:insertrapq_insert_edge}} can construct the actual path (Line \ref{line:expand_addresult}) and return it as a part of the resulting sgt, i.e., it populates the implicit payload attribute $\mathcal{D}$ of the resulting sgt with the sequence of edges that forms the resulting path.
The cost of this operation is $\mathcal{O}(l)$ where $l$ is the length of the resulting path.

$\Delta-\texttt{PATH}$ guarantees that the expiry timestamp of a node $(u,s)$ in $T_x$ is equal to largest expiry timestamp of all paths between $x$ and $u$ in the snapshot graph with a label $l$ such that $s = \delta^*(s_0, l)$.
Consequently, for a node $(u,s) \in T_x$  with expiry timestamp smaller than $t$, there cannot be another path from $x$ to $u$ with an equivalent label that is valid at time $t$.
Consider the spanning tree given in Figure \ref{fig:tree-v2-new}.
Algorithm \textbf{\ref{alg:arbitrary}} can directly determine, without additional processing, that nodes $(z,1)$ and $(t,1)$ are expired as their expiry timestamp is 31.
Thus, at any given time $t$, Algorithm \textbf{\ref{alg:arbitrary}} can simply ignore a node $(u,s) \in T_x$  with expiry timestamps smaller than $t$ (Line \ref{line:insertrapq_updateexpiry}) and such nodes can be removed from $\Delta-\texttt{PATH}$.
To prevent $\Delta-\texttt{PATH}$ from growing unboundedly due to expired tuples, a background process periodically purges expired tuples from $\Delta-\texttt{PATH}$.

The following example illustrates how the \textit{negative tuple} and \textit{direct} approaches differ, respectively for \cite{Pacaci_2020} and our proposed algorithm.

\begin{example}\label{ex:no-expiry}
Consider the same real-time notification query as in Example \ref{ex:navigation_algorithm}.
Both approaches behave similarly until $t=28$ as all vertex-state pairs in $T_x$ have a single derivation at $t=27$ (Figure \ref{fig:tree-v1}).
Upon arrival of the sgt $(y,u,HI,[28,37), \mathcal{D} = \{(y,HI,u)\})$ at $t=28$, the negative tuple approach as in \cite{Pacaci_2020} does not update $T_x$ as $(u,1)$ is already in $T_x$, whereas the direct approach as used in this paper updates the validity interval and the parent pointer of $(u,1) \in T_x$ (Line \ref{line:insertrapq_updateexpiry} in Algorithm \textbf{\ref{alg:arbitrary}}).
Then, incoming sgts at times $t=28$ and $t=29$ are processed similarly, adding $(v,1)$ and $(s,1)$ as children of $(u,1)$.
Figures \ref{fig:tree-v2-new} and \ref{fig:tree-v2-old} depict the corresponding spanning trees at $t=30$ for the \textit{direct} and the \textit{negative tuple} approaches, respectively.
Note that in Figure \ref{fig:tree-v2-new}, the validity intervals of nodes in the subtree rooted at node $(u,1)$ reflects the newly discovered path from $x$ to $u$ through $y$ in $G_{30}$.
The \textit{negative tuple} and the \textit{direct} approach differs at $t=31$ as multiple nodes expire. 
The \textit{negative tuple} used as in \cite{Pacaci_2020} marks the entire subtree of $(z,1)$ as potentially expired (Figure \ref{fig:tree-v2-old}), and performs a traversal of the snapshot graph $G_{31}$ to find alternative, valid paths for expired nodes.
These traversals undo the effect of expired sgts via explicit deletions.
Upon discovering alternative paths for nodes $(u,1), (v,1)$ and $(s,1)$ that are valid at time $t=31$, they are re-inserted into $T_x$.
Instead, our proposed algorithm can \textit{directly} determine the expired nodes based on the validity intervals (nodes $(z,1)$ and $(t,1)$ as shown in Figure \ref{fig:tree-v2-new}) without additional processing.
\end{example}

\subsubsection{Explicit Deletions}
\label{sec:explicit-deletions}

Physical implementations of \textsf{PATTERN} and \textsf{\small PATH} rely on the \textit{direct} approach that utilizes the temporal pattern of sliding windows for state maintenance; the expiration timestamps are used to \textit{directly} locate expired tuples.
On append-only streaming graphs, existing sgts only expire due to window movements.
Albeit rare, certain applications might require explicit deletions of previously inserted sgts, which necessitates the use of the \textit{negative tuple} approach to handle such explicit deletions.\footnote{The \textit{negative tuple} approach  can also be used to signal expirations through an explicit deletion of the corresponding sgt to undo its effect \cite{ghanem2006incremental, golabo05, Pacaci_2020}. 
Indeed, the negative-tuple approach can be used for incremental evaluation of an arbitrary computation, whereas direct approach is applicable to negation-free queries over append-only streaming graphs \cite{golabo05}.}
Deleted sgt, with an additional flag to denote deletion -- i.e., a negative tuple, is used to undo the effect of the original sgt on the operator state and to invalidate previously reported results, if necessary.
For pipelined hash join, which is used in the implementation of \textsf{PATTERN}, processing of negative tuples is the same as original input tuples: a negative tuple is removed from its corresponding hash table and other tables are probed to find corresponding deleted results.


The use of negative tuples for explicit deletions in the context of RPQ evaluation is first proposed by Pacaci et al \cite{Pacaci_2020}.
Here, we describe how to adopt the \textit{negative tuple} approach for explicit deletions of sgts in Algorithm \textbf{\ref{alg:arbitrary}}.
In brief, upon explicit deletion of an sgt, we first identify tree-edges that disconnect spanning trees in $\Delta-\texttt{PATH}$. 
For each such tree-edge, we first mark the nodes in the subtree that are disconnected due to the explicit deletion.
Then, for each node in this subtree, we use a Dijkstra-based traversal over the snapshot graph to find an alternative path with the largest expiry timestamp.
Dijkstra's algorithm guarantees that we can efficiently find the path with the largest expiry timestamp for each marked node, consistent with Definition \ref{def:delta}.
A marked node is removed from the spanning tree only if there is no alternative valid path.
Deletion of a non-tree edge does not require any modification as it leaves spanning trees unchanged.


\section{Experimental Analysis}
\label{sec:experiments}

Our objective is to demonstrate the feasibility of implementing a performant system that incorporates the algebraic framework we propose in this paper.
Using the prototype implementation described in \S \ref{sec:implementation}, we first provide an end-to-end performance analysis of our algebraic approach for persistent evaluation of streaming graph queries (\S \ref{sec:experiments-query-processing}).
Then, we assess the scalability by varying the window size $\mathcal{T}$ and the slide interval $\beta$ (\S \ref{sec:experiments-scalability}).
Finally, we highlight the benefits of the proposed SGA in exploring the rich plan space through transformation rules and demonstrate the potential performance improvements of exploring this plan space (\S \ref{sec:experiments-transformation}).

\subsection{Experimental Setup}
\label{sec:experiments-setup}

\subsubsection{Setup}
Experiments are run on a Linux server with 32 physical cores and 256GB memory.
For each query and configuration, we report the tail latency of each
window slide, i.e., the total time to process all arriving and expired sgts upon window movement and to produce new results, and the average throughput after ten minutes of processing on warm caches.

\subsubsection{Datasets}

We use \textbf{Stackoverflow} (SO) and \textbf{LDBC SNB} (SNB) graphs for our experimental analysis;
these are publicly available, large-scale graphs with labelled and timestamped edges on which persistent queries with complex graph patterns can be formulated.
SO is a temporal graph of user interactions on the stackoverflow website containing 63M interactions (edges) of 2.2M users (vertices), spanning 8 years \cite{paranjape2017motifs}, and SNB is a synthetic social network graph that simulates the interactions of an online social network \cite{sigmod15_erling:2015}.
We extract the update stream of the LDBC workload that contains 8 different types of interactions, and we use \textsf{\small replyOf}, \textsf{\small hasCreator} and \textsf{\small likes} edges between users and posts, and \textsf{\small knows} edges between users.
We use a scale factor of 10 with 7.2M users and posts (vertices) and  40M user interactions (edges).
SO contains only a single type of vertex and 3 different edge labels, and its cyclic nature causes a high number of intermediate results and resulting paths; so it is the most challenging one for the proposed algorithms.
We set the window size $\mathcal{T}$ to 1 month and the slide interval $\beta$ to 1 day unless specified otherwise.

\subsubsection{Workloads}

To the best of our knowledge, no current benchmark exists featuring RQ for graph DBMSs. 
The existing benchmarks are limited to 
UCRPQ,  
thus not capturing the full expressivity of RQ even for static graphs. 
Streaming RDF benchmarks such as LSBench (\url{https://code.google.com/archive/p/lsbench/})  and Stream WatDiv ~\cite{gao:2018aa} only focus on SPARQL v1.0 (thus not even including simple RPQs), and their workloads do not contain any recursive queries.
Hence, we formulate a set SGQ from existing UCRPQ-based workloads as follows:
we collect a set of graph patterns in the form of UCRPQ from existing benchmarks and studies \cite{yakovets2016query, bonifati2019navigating, Pacaci_2020, sigmod15_erling:2015, bagan2016gmark}, and we compose a set of complex graph patterns from those by applying a Kleene star over each graph pattern.
Table \ref{tab:query_shapes} lists the set of graph patterns of increasing expressivity (from RPQ to complex RQ with complex graph patterns) that we use to define streaming graph queries.
$Q_1 - Q_4$ are commonly used RPQs in existing studies \cite{yakovets2016query, bonifati2019navigating, Pacaci_2020}, and we use those to test our \textsf{\small PATH} operator. 
$Q_5$ \& $Q_6$ are CRPQ-based complex graph patterns based on  SNB queries \textsf{\small IS7} and \textsf{\small IC7} \cite{sigmod15_erling:2015}.
For instance, $Q_6$ -- \textsf{\small IC7} of SNB -- with edge labels \textsf{\small knows}, \textsf{\small likes} and \textsf{\small hasCreator} asks for \textsf{\small recent liker}s of a person's messages that are also connected by a path of friends.
$Q_7$ -- Ex. \ref{ex:alert} -- is the most expressive RQ-based complex graph patterns we use to demonstrate the abilities of the proposed SGA to unify subgraph pattern and path navigation queries in a structured manner and to treat paths as first-class citizens.
It defines a path query over the complex graph pattern of $Q_6$; it finds arbitrary length paths where users are connected by the \textsf{\small recentLiker} pattern.
Note that this query cannot be expressed in existing graph query languages such as Cypher and SPARQL.
Finally, for each dataset, we instantiate the query workload from these graph patterns by choosing appropriate predicates, i.e., edge labels, for each query edge and by setting the duration of time-based sliding windows $\mathcal{W}_{\mathcal{T}}$ as described above. 

\begin{table}
\caption{ 
$Q_1 - Q_4$ correspond to common RPQ observed in real-world query logs \cite{bonifati2019navigating}, and $Q_5 - Q_7$ are Datalog encodings of RQ-based complex graph patterns that we use to define streaming graph queries.
$Q_5$ and $Q_6$ correspond to complex graph patterns of LDBC SNB queries $IS7$ and $IC7$ \cite{sigmod15_erling:2015}, respectively, and $Q_7$ corresponds to the complex graph pattern given in Example \ref{ex:alert} that is defined as a recursive path query over the graph pattern of $Q_6$. $a,b $ and $c$ correspond to edge predicates that are instantiated based on the dataset characteristics. 
}
\small
	\centering
		\begin{tabular}{ | r | l | }
	\hline
        Name & Query \\
	\hline
	$Q_1$ & $?x,?y \ \leftarrow \ ?x\ a^* \ ?y$  \\ 
	$Q_2$ & $?x,?y \ \leftarrow \ ?x\ a \circ b^* \ ?y$  \\
	$Q_3$ & $?x,?y \ \leftarrow \ ?x\ a \circ b^*$  $\circ c^* \ ?y$ \\
	$Q_4$ & $?x,?y \ \leftarrow \ ?x\ (a \circ b \circ c)^+ \ ?y$ \\
    $Q_5$ & $RR(m1,m2) \ \leftarrow \ a(x,y),\ b(m1, x),\ b(m2, y),\ c(m2, m1)$ \\
    $Q_6$ & $RL(x,y) \ \leftarrow \ a^+(x,y),\ b(x, m),\ c(m, y)$ \\
    \multirow{2}{*}{$Q_7$} & $RL(x,y) \ \leftarrow \ a^+(x,y),\ b(x, m),\ c(m, y)$ \\
    & $Ans(x, m) \ \leftarrow \ RL^+(x,y), c(m, y)\ $ \\
	\hline
	\end{tabular}
\label{tab:query_shapes}
\end{table}

\subsection{Query Processing Performance}
\label{sec:experiments-query-processing}

\subsubsection{Throughput \& Tail Latency}

\begin{table*}
\caption{(Tput) The throughput (edges/s) and (TL) the tail latency (s) of SGA and DD for queries in Table \ref{tab:query_shapes} on SO and SNB graphs with $|W| =$ 30 days and $\beta = $ 1 day. }
\small
    \centering
    \begin{tabular}{|c c|c c|c c|c c|c c|c c|c c|c c|c c|}
    \hline
        \multirow{2}{*}{Graph} & \multirow{2}{*}{System} & \multicolumn{2}{c|}{$Q_1$} & \multicolumn{2}{c|}{$Q_2$} &  \multicolumn{2}{c|}{$Q_3$} & \multicolumn{2}{c|}{$Q_4$} & \multicolumn{2}{c|}{$Q_5$} & \multicolumn{2}{c|}{$Q_6$} & \multicolumn{2}{c|}{$Q_7$} \\ 
        & & Tput & TL & Tput & TL & Tput & TL & Tput & TL & Tput & TL & Tput & TL & Tput & TL \\
    \hline
    \hline
         \multirow{2}{*}{SO} & SGA & \textbf{2762} & \textbf{3.3} & \textbf{8513} & \textbf{4.3} & \textbf{413} & \textbf{120} & \textbf{379} & 102.4 & \textbf{231064} & \textbf{0.3} & \textbf{374} & \textbf{52.7} & \textbf{376} & \textbf{56.3} \\
         & DD & 1209 & 6.3 & 4512 & 5.8 & 368 & 121.7 & 374 & \textbf{82.8} & 63330 & 1 & 283 & 72.6 & 275 & 74 \\
    \hline
        \multirow{2}{*}{SNB} & SGA & 97187 & 1.5 & 237313 & 1.9 & 245766 & 1.9 & 277475 & 0.4 & \textbf{13345} & \textbf{79.1} & \textbf{428592} & \textbf{0.8} & \textbf{131250} & \textbf{10.2} \\
         & DD & \textbf{121133} & \textbf{0.8} & \textbf{299245} & \textbf{1.2} & \textbf{316267} & \textbf{1.1} & \textbf{303068} & \textbf{0.2} & 12053 & 109.5 & 402048 & 0.9 & 21284 & 141 \\
    \hline
    \end{tabular}
    \label{tab:tput-latency}
\end{table*}

Table \ref{tab:tput-latency} (SGA) shows the aggregated throughput and tail latency of our streaming graph query processor for all queries in Table \ref{tab:query_shapes}.
We discard each streaming graph edge whose label is not in a given SGQ.
Tail latencies reflect the $99^{th}$ percentile latency of processing a window slide and produce the corresponding resulting sgts.
Across queries, the performance is lower for SO graph because it is dense and cyclic.
The throughput  ranges from hundreds of edges-per-second for the SO to hundreds of thousands of edges-per-second for SNB.

\subsubsection{Comparative Analysis}
\label{sec:experiments_comparative}

Existing work on query processing over streaming data such as data stream management systems and streaming RDF systems cannot process queries in Table \ref{tab:query_shapes} as they focus on relational queries and SPARQL v1.0, respectively (\S \ref{sec:related}).
To the best our knowledge, TD with its DD layer is the only general-purpose system that can be used to incrementally evaluate recursive computations that are modelled as cyclic dataflows.
Table \ref{tab:tput-latency} (DD) reports the throughput and tail latency of DD dataflows for all queries in Table \ref{tab:query_shapes}.
Overall, our SGA-based query processor outperforms the DD baseline on SO and provides a competitive performance on the  SNB dataset. On SNB, $Q_6$ \& $Q_7$ do not have the Kleene-plus over  $a$ as it causes DD to timeout.
Due to highly cyclic structure of SO, there are many alternative paths between each pair of vertices, and our streaming RPQ algorithm for \textsf{\small PATH} implementation  maintains a compact representation of valid path segments and utilizes the temporal patterns of sliding window movements to simplify expirations (\S \ref{sec:physical-algebra}).
DD-based query processor provides better performance on linear path queries $Q_1$--$Q_4$ on SNB, but not others. 
This is due to the tree-shaped structure of \textsf{\small replyOf} edges in SNB, where there is only one path between a pair of vertices, so \textsf{\small PATH} specific optimizations do not apply. 
Performance variations on SNB suggest optimization opportunities for recursive graph queries when selecting physical operator implementations, as in the case for streaming relational joins \cite{golabo05}.
These results demonstrate the feasibility of our algebraic approach for evaluating SGQ and our physical operator implementations.

\subsection{Sensitivity Analysis}
\label{sec:experiments-scalability}

\begin{figure*}
\centering
    \begin{subfigure}[t]{0.45\textwidth}
        \includegraphics[width=\textwidth]{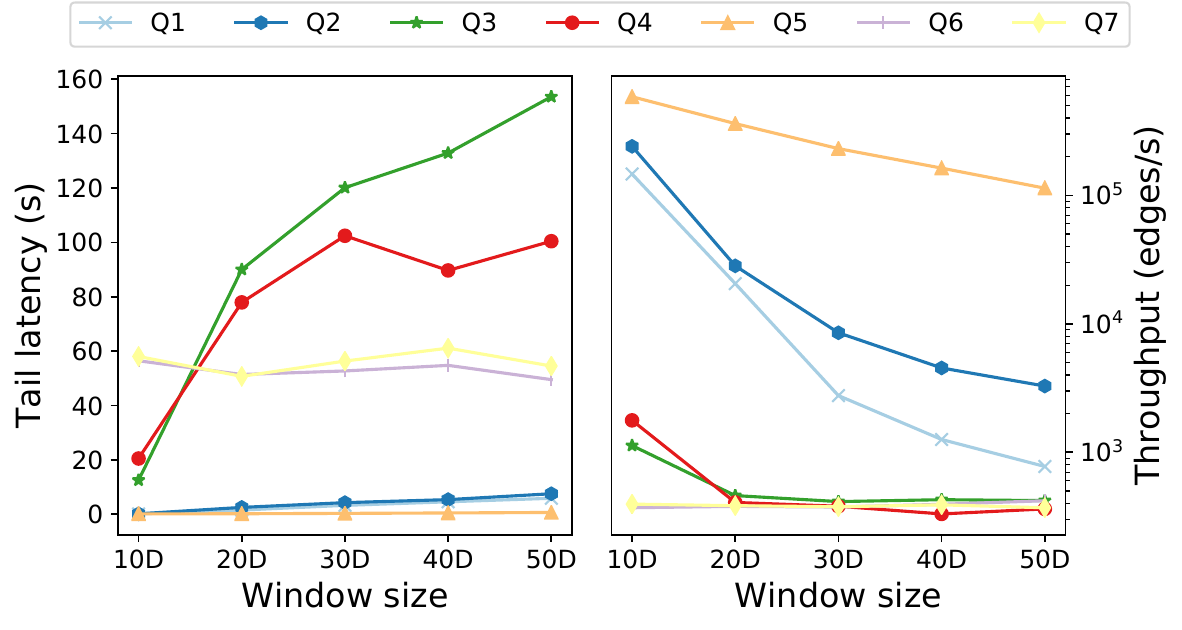}
        \caption{Window Size}
        \label{fig:so-scalability-window}
    \end{subfigure}
    \begin{subfigure}[t]{0.45\textwidth}
        \includegraphics[width=\textwidth]{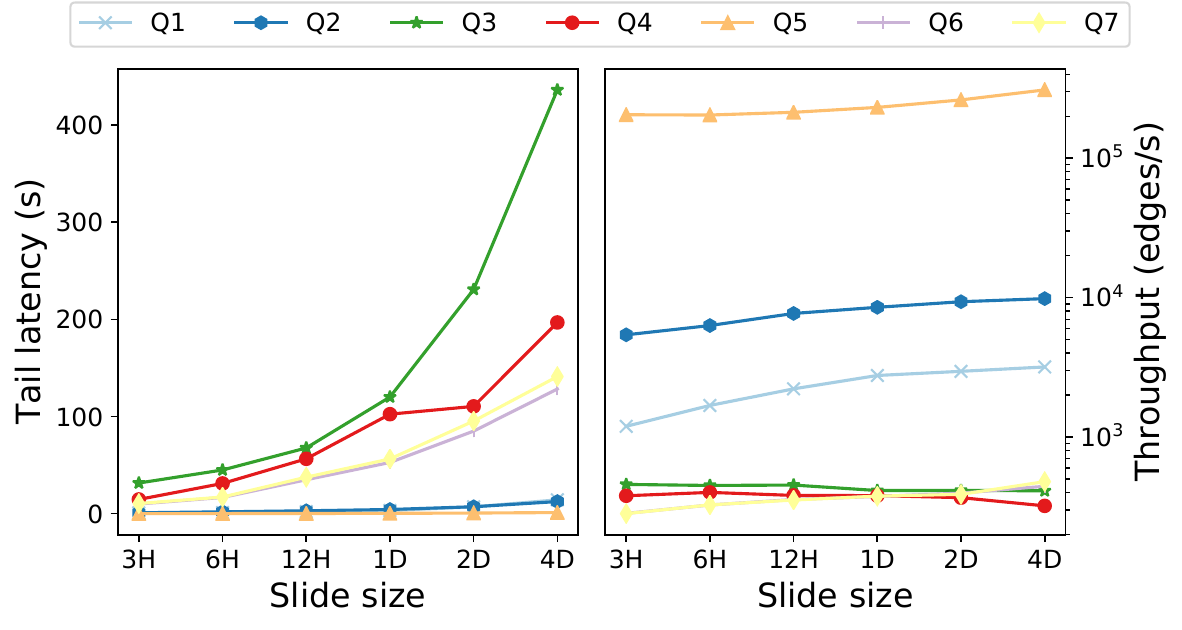}
        \caption{Slide}
        \label{fig:so-scalability-slide}
    \end{subfigure}
    \caption{The tail latency of each window slide and the aggregate throughput of the proposed streaming graph query processor with increasing (a) window size $\mathcal{T}$ and (b) slide interval $\beta$ on SO graph.} 
    \label{fig:so-scalability}
\end{figure*}

\begin{figure}
\centering
    \includegraphics[width=\linewidth]{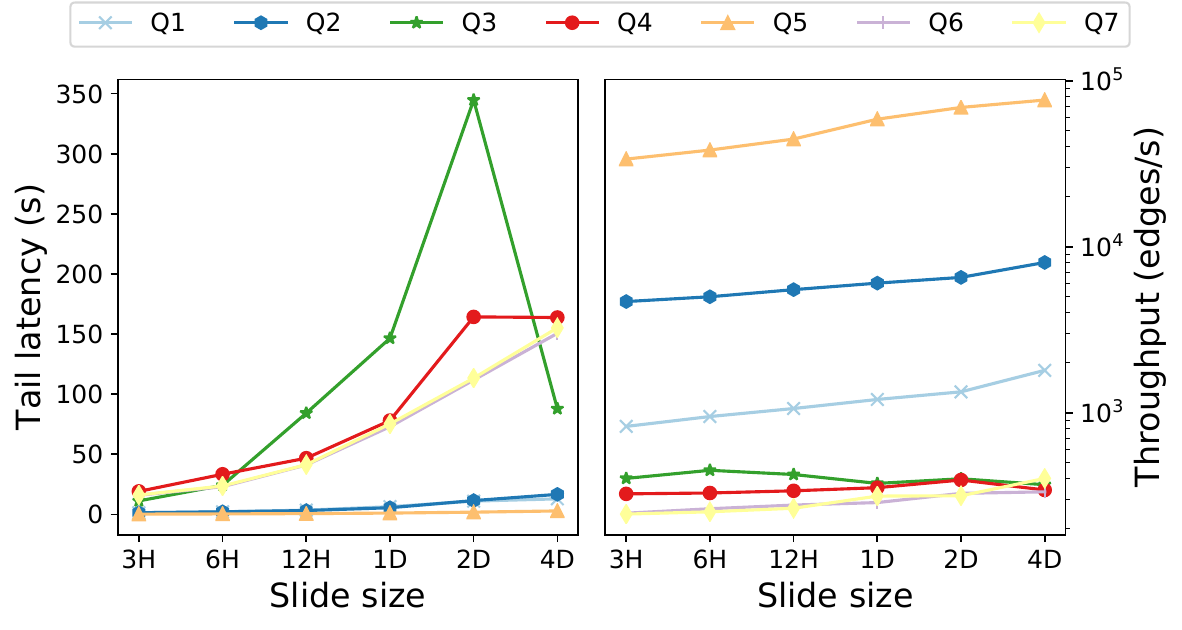}
    \caption{The tail latency of each window slide and the aggregate throughput of SGQ evaluation on DD with increasing slide interval $\beta$ on SO graph.}
    \label{fig:so-scalability-slide-dd}
\end{figure}

In this section, we analyze the impact of the window size $\mathcal{T}$ and the slide interval $\beta$ on end-to-end query performance of the proposed streaming graph query processor.
We use SO graph for this experiment as its dense, cyclic structure stresses our operator implementations.
Fig. \ref{fig:so-scalability-window}  reports the aggregate throughput and the tail latency for each query across various window sizes.
As expected, the throughput of all tested queries decreases with increasing $\mathcal{T}$, as a larger window size increases the \# of sgts in each window.
Similarly, the tail latency of each window slide increases with the increasing window size.

We also assess the impact of the slide interval $\beta$ on performance.
As previously mentioned, the slide interval $\beta$ controls the time-granularity at which the sliding window progresses, and our prototype implementation uses $\beta$ to control the input batch size.
Figure \ref{fig:so-scalability-slide} shows that the aggregate throughput and the tail latency for each query remain stable across varying slide intervals. 
This is due to tuple-oriented implementation of physical operators of SGA; SGA operators are designed to process each incoming tuple eagerly in favour of minimizing tuple-processing latency, and they do not utilize batching to improve throughput with larger batch sizes.
Consequently, the tail latency of window movements increases with increasing slide interval.
This is in contrast to DD whose throughput increases with increasing $\beta$ as shown in Figure \ref{fig:so-scalability-slide-dd}.
DD and its underlying indexing mechanism, i.e., shared arrangements \cite{mcherry_shared_vldb20}, are designed to utilize batching and improve throughput with increasing batching size: all sgts that arrive within one interval are batched together with a single logical timestamp (epoch) and DD operators can explore the latency vs throughput trade-off by changing the granularity of each epoch. 
The investigation of batching within SGA operators and the identification of other optimization opportunities is a topic of future work.





\subsection{Exploring the Plan Space}
\label{sec:experiments-transformation}


\begin{figure}
\centering
    \includegraphics[width=0.8\linewidth]{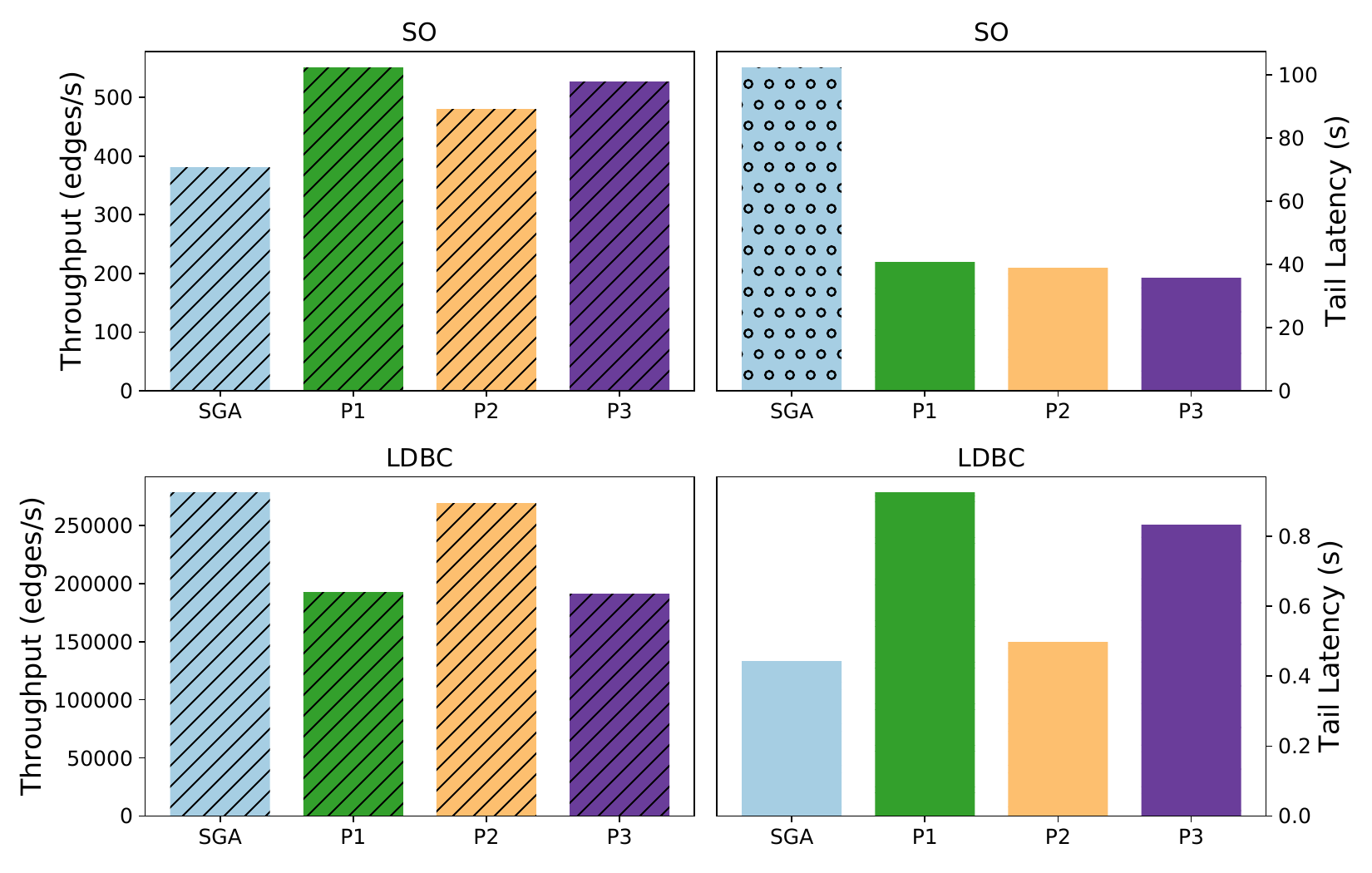}
    \caption{The throughput and tail latency of (a) $Q_4$ on (top) SO and (bottom) SNB for the default SGA plan and an alternative equivalent physical plan generated via SGA transformation rules (Section \ref{sec:query-transformation}).}
    \label{fig:query4-plan}
\end{figure}

\begin{figure}
\centering
    \includegraphics[width=0.8\linewidth]{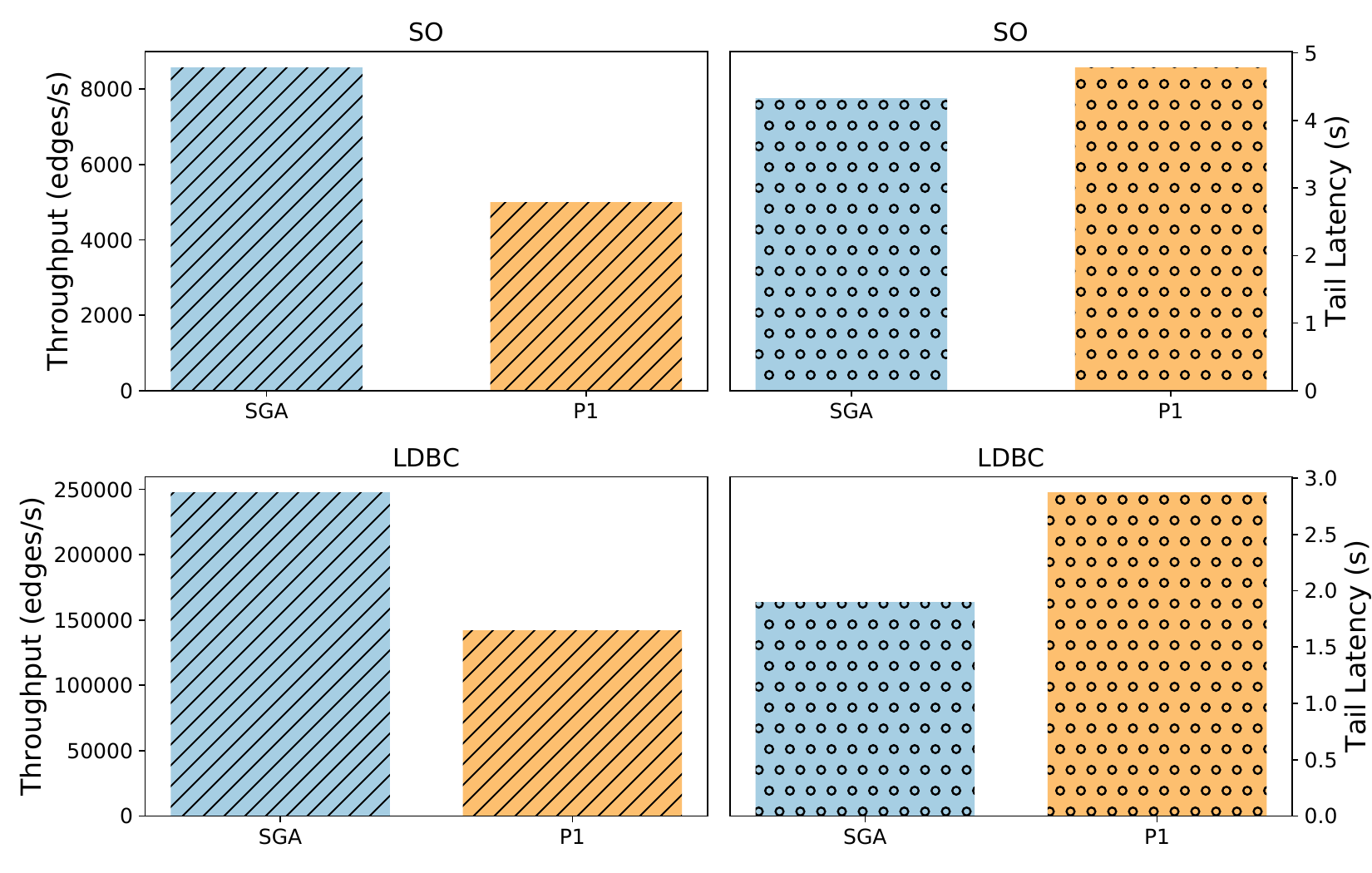}
    \caption{The throughput and tail latency of (a) $Q_2$ on (top) SO and (bottom) SNB for the default SGA plan and an alternative equivalent physical plan generated via SGA transformation rules (Section \ref{sec:query-transformation}).}
    \label{fig:query2-plan}
\end{figure}

\begin{figure}
\centering
    \includegraphics[width=0.8\linewidth]{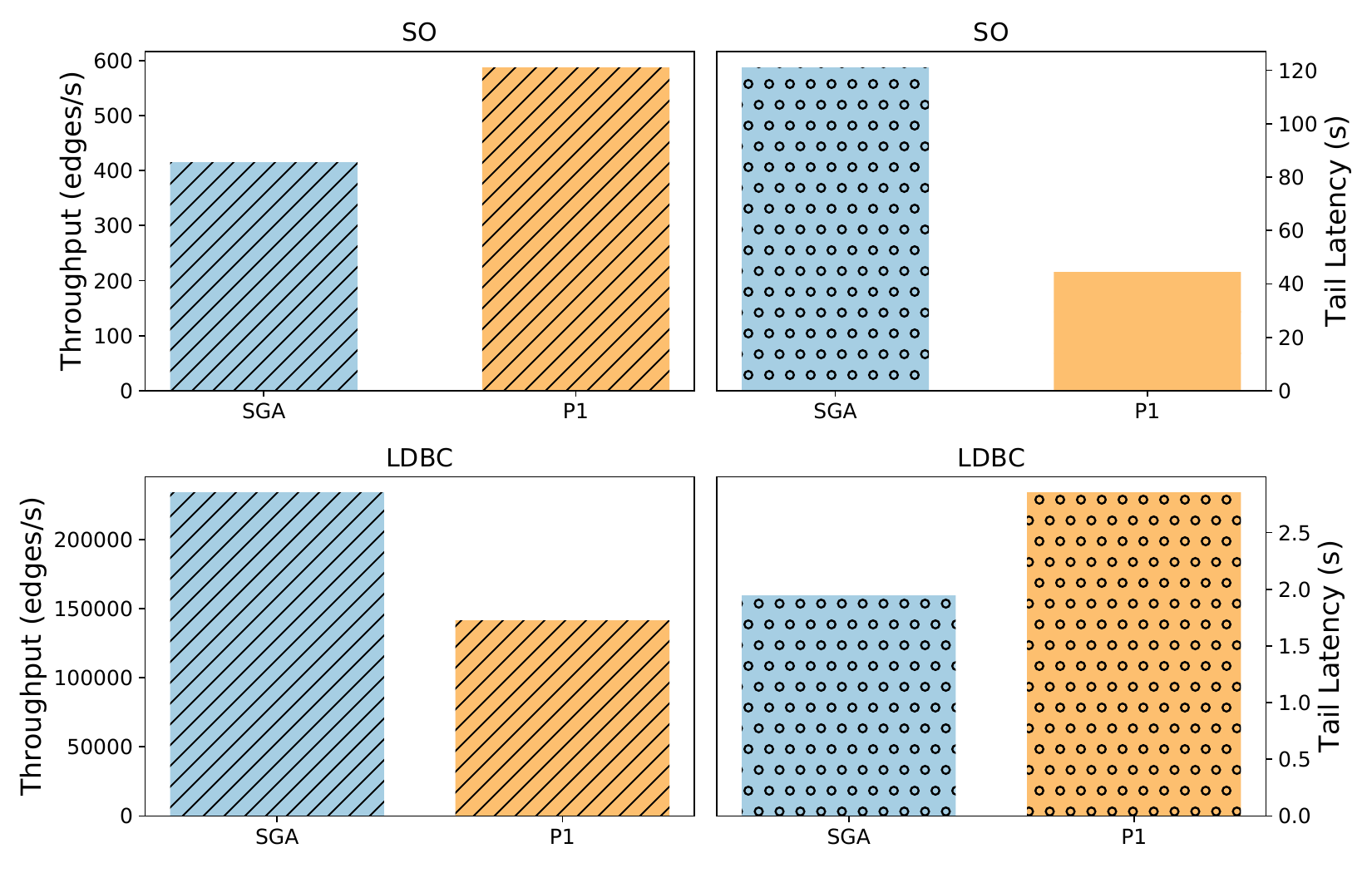}
    \caption{The throughput and tail latency of (a) $Q_3$ on (top) SO and (bottom) SNB for the default SGA plan and an alternative equivalent physical plan generated via SGA transformation rules (Section \ref{sec:query-transformation}).}
    \label{fig:query3-plan}
\end{figure}


SGA proposed in this paper enables a rich foundation for logical SGQ optimization through query rewriting as previously discussed (\S \ref{sec:query-transformation}). In this paper, we do not address the full scope of optimization issues (full optimizer development is the topic of ongoing research), but 
we design a micro-benchmark to highlight the possibilities provided by SGA. 
In particular, we choose $Q_4$ (Table \ref{tab:query_shapes}) as its linear pattern combined with a Kleene plus demonstrates the potential benefits of SGA transformation rules involving SGA's novel \textsf{PATH} operator.
Fig. \ref{fig:query4-plan} demonstrates the throughput and the tail latency of different plans obtained from following equivalent SGA expressions for $Q_4$:

\begin{itemize}
    \item SGA: $\mathcal{P}_{d^+}^l(\Join_{trg_1 = src_2 \land trg_2 = src_3}^{src_1, trg_3, d}( S_a, S_b, S_c))$
	\item P1: $\mathcal{P}_{(a \cdot b \cdot c)^+}^l(S_a, S_b, S_c)$
    \item P2: $\mathcal{P}_{(a \cdot d)^+}^l( S_a, \Join_{trg_1 = src_2}^{src_1, trg_2, d}(S_b, S_c))$
    \item P3: $\mathcal{P}_{(d \cdot c)^+}^l( S_c, \Join_{trg_1 = src_2}^{src_1, trg_2, d}(S_a, S_b))$
\end{itemize}

The first expression, \textbf{SGA}, is the canonical SGA expression for $Q_4$ that is generated by the Algorithm \textbf{\ref{alg:sga-generator}}.
Such plans are called \textit{loop-caching} in literature as they enable re-use of the intermediate results for the base pattern $(a\cdot b\cdot c)$ \cite{yakovets2016query}\footnote{It is also the only plan available to DD baseline that performs a fixed-point iteration over the base pattern $(a\cdot b\cdot c)$.}.
\textbf{P1}, \textbf{P2} and \textbf{P3} are obtained from the canonical SGA expression using the transformation rules given in \S \ref{sec:query-transformation}, and represent novel plans that are possible due to novel \textsf{\small PATH} operator.
Fig. \ref{fig:query4-plan} clearly illustrates the potential benefits of exploring the rich plan space offered by SGA: some of the newly computed plans provide up to 60\% increase in throughput and 60\% reduction in the latency. 
We observe a similar behaviour on other path queries $Q_2$ and $Q_3$ (up to 50\% difference in throughput).
These results suggest further optimization opportunities for logical query optimization as query rewrites that are generated by SGA transformation rules can provide significant performance benefits for evaluating SGQ over streaming graphs.

\subsection{Alternative Physical Operators}
\label{sec:experiments-alternative}

\begin{table*}
\caption{ The impact of  \textbf{\ref{alg:arbitrary}} on the performance of the throughput (edges/s) and (TL) the tail latency (s) of the queries in Table \ref{tab:query_shapes} on SO and SNB graphs with $|W| =$ 30 days and $\beta = $ 1 day. }
\small
    \centering
    \begin{tabular}{|c c|c c|c c|c c|c c|c c|c c|c c|c c|}
    \hline
        \multirow{2}{*}{Graph} & \multirow{2}{*}{System} & \multicolumn{2}{c|}{$Q_1$} & \multicolumn{2}{c|}{$Q_2$} &  \multicolumn{2}{c|}{$Q_3$} & \multicolumn{2}{c|}{$Q_4$} & \multicolumn{2}{c|}{$Q_5$} & \multicolumn{2}{c|}{$Q_6$} & \multicolumn{2}{c|}{$Q_7$} \\ 
        & & Tput & TL & Tput & TL & Tput & TL & Tput & TL & Tput & TL & Tput & TL & Tput & TL \\
    \hline
    \hline
         \multirow{2}{*}{SO} &\textbf{\ref{alg:arbitrary}} & \textbf{2884} & 4 & \textbf{9074} & 4.9 & 391 & 177 & 348 & \textbf{94.9} & \textbf{234058} & 0.4 & \textbf{625} & \textbf{51.4} & 353 & \textbf{52.6}  \\
         
         & Improvement &4.4\% &	-21.2\% &	6.5\% &	-13.9\% &	-5.3\% &	-47.5 \% &	-8.1\% &	7.3\% &	1.2\% &	-33.3 \% &	67.1\% &	2.4\% &	-6.1\% &	6.5\%  \\
    \hline
        \multirow{2}{*}{SNB}  & \textbf{\ref{alg:arbitrary}} & 95903 &\textbf{ 1.4} & \textbf{244653} & \textbf{1.8} & 224342 & 1.9 & \textbf{278647} & 0.4 & \textbf{14000} & 79.5 & \textbf{450957} & 0.8 & 130651 & 10.8 \\
        
        & Improvement & -1.3\% &	6.6\% &	3.1 \% &	5.2\% &	-8.7\% &	- &	0.4\% &	- &	4.9\% &	-0.5\% &	5.2\% &	- &	-0.4\% &	-5.8\%  \\
        
    \hline
    \end{tabular}
    \label{tab:rpq-comparison}
\end{table*}

Our prototype implementation employs the automata-based streaming RPQ algorithm of \cite{Pacaci_2020} as the physical implementation of the \textsf{\small PATH} operator.
Section \ref{sec:path-navigation-implementation} describes an alternative physical implementation (\textit{Streaming Path Navigation -- \textbf{\ref{alg:arbitrary}}}) for the novel \textsf{\small PATH} operator of our proposed SGA based on the \textit{direct approach}.
Table \ref{tab:rpq-comparison} compares its impact on the performance of our proposed streaming graph query processor.
The performance impact of adopting the Algorithm \textbf{\ref{alg:arbitrary}} for \textsf{\small PATH} varies depending on the workload characteristics: it provides improvements in throughput for most queries on the SO graph, whereas the performance differences on the SNB graph is small.
This can be attributed to the cyclic nature of the SO graph.
A higher number of intermediate path segments increases the size of the internal operator state that the physical implementations of the \textsf{\small PATH} operator need to maintain.
Overall, these results illustrate the potential benefits of exploring alternative operator implementations. 
In particular, cost-based optimization techniques for determining suitable operator implementations based on data and query characteristics might provide significant performance benefits for evaluating SGQ over streaming graphs.

\section{Conclusion and Future Work}
\label{sec:conclusion}

This paper introduces a general-purpose query processing framework for streaming graphs that consists of (i) streaming graph query model and algebra with well-founded semantics, and 
(ii) a prototype streaming graph query processor as an embodiment of the proposed framework.
The SGQ model addresses the requirements we set forth for querying streaming graphs by capturing both common features of existing graph query languages and novel features that are beyond the power of existing languages (such as the treatment of paths as first-class citizens, powerful path patterns and composability).
SGA and its transformation rules provide the foundational framework to express and evaluate streaming graph queries.
Our prototype streaming graph query processor exemplifies physical operator implementations specific to streaming graph queries due to this framework.
Experimental analyses on real-world and synthetic streaming graphs demonstrate the feasibility and the potential performance gains of our approach. 

SGQ and SGA establish the basis of systematic study of query processing issues over streaming graphs. We are now engaged in research along two dimensions: (i)  designing an SGA-based query optimizer for the systematic exploration of the rich plan space using  SGA’s transformation rules,
and (ii)  incorporating attribute-based predicates to fully support the property graph model.

Additional work that would enrich the implementation include the development of alternative physical operators and additional transformation rules for plan space enumeration. Given that the SGQ-SGA framework is complete with a clear semantics (as captured in RQ model), and given that SGQ incorporates the capabilities of existing and emerging query languages, future work can focus on efficient implementation.

\color{black}



\bibliographystyle{ACM-Reference-Format}

\balance 

\bibliography{publishers,publications,references_sga_icde}


\begin{thebibliography}{74}


\ifx \showCODEN    \undefined \def \showCODEN     #1{\unskip}     \fi
\ifx \showDOI      \undefined \def \showDOI       #1{#1}\fi
\ifx \showISBNx    \undefined \def \showISBNx     #1{\unskip}     \fi
\ifx \showISBNxiii \undefined \def \showISBNxiii  #1{\unskip}     \fi
\ifx \showISSN     \undefined \def \showISSN      #1{\unskip}     \fi
\ifx \showLCCN     \undefined \def \showLCCN      #1{\unskip}     \fi
\ifx \shownote     \undefined \def \shownote      #1{#1}          \fi
\ifx \showarticletitle \undefined \def \showarticletitle #1{#1}   \fi
\ifx \showURL      \undefined \def \showURL       {\relax}        \fi
\providecommand\bibfield[2]{#2}
\providecommand\bibinfo[2]{#2}
\providecommand\natexlab[1]{#1}
\providecommand\showeprint[2][]{arXiv:#2}

\bibitem[\protect\citeauthoryear{Abadi, Ahmad, Balazinska, \c{C}etintemel,
  Cherniack, Hwang, Lindner, Maskey, Rasin, Ryvkina, Tatbul, Xing, and
  Zdonik}{Abadi et~al\mbox{.}}{2005}]%
        {cidr05_abadiabcchlmrrtxz05}
\bibfield{author}{\bibinfo{person}{Daniel~J. Abadi}, \bibinfo{person}{Yanif
  Ahmad}, \bibinfo{person}{Magdalena Balazinska}, \bibinfo{person}{Ugur
  \c{C}etintemel}, \bibinfo{person}{Mitch Cherniack},
  \bibinfo{person}{Jeong-Hyon Hwang}, \bibinfo{person}{Wolfgang Lindner},
  \bibinfo{person}{Anurag Maskey}, \bibinfo{person}{Alex Rasin},
  \bibinfo{person}{Esther Ryvkina}, \bibinfo{person}{Nesime Tatbul},
  \bibinfo{person}{Ying Xing}, {and} \bibinfo{person}{Stanley~B. Zdonik}.}
  \bibinfo{year}{2005}\natexlab{}.
\newblock \showarticletitle{The Design of the Borealis Stream Processing
  Engine}. In \bibinfo{booktitle}{\emph{Proc. 2nd Biennial Conf. on Innovative
  Data Systems Research}}. \bibinfo{pages}{277--289}.
\newblock


\bibitem[\protect\citeauthoryear{Abadi, Carney, {\c C}etintemel, Cherniack,
  Convey, Lee, Stonebraker, Tatbul, and Zdonik}{Abadi et~al\mbox{.}}{2003}]%
        {abadi2003}
\bibfield{author}{\bibinfo{person}{Daniel~J. Abadi}, \bibinfo{person}{Don
  Carney}, \bibinfo{person}{Ugur {\c C}etintemel}, \bibinfo{person}{Mitch
  Cherniack}, \bibinfo{person}{Christian Convey}, \bibinfo{person}{Sangdon
  Lee}, \bibinfo{person}{Michael Stonebraker}, \bibinfo{person}{Nesime Tatbul},
  {and} \bibinfo{person}{Stan Zdonik}.} \bibinfo{year}{2003}\natexlab{}.
\newblock \showarticletitle{Aurora: a new model and architecture for data
  stream management}.
\newblock \bibinfo{journal}{\emph{VLDB J.}} \bibinfo{volume}{12},
  \bibinfo{number}{2} (\bibinfo{year}{2003}), \bibinfo{pages}{120--139}.
\newblock
\showISSN{1066-8888}


\bibitem[\protect\citeauthoryear{Abiteboul, Hull, and Vianu}{Abiteboul
  et~al\mbox{.}}{1995}]%
        {abiteboul1995foundations}
\bibfield{author}{\bibinfo{person}{Serge Abiteboul}, \bibinfo{person}{Richard
  Hull}, {and} \bibinfo{person}{Victor Vianu}.}
  \bibinfo{year}{1995}\natexlab{}.
\newblock \bibinfo{booktitle}{\emph{Foundations of databases}}.
  Vol.~\bibinfo{volume}{8}.
\newblock \bibinfo{publisher}{Addison Wesley}.
\newblock


\bibitem[\protect\citeauthoryear{Aghasadeghi, Moffitt, Schelter, and
  Stoyanovich}{Aghasadeghi et~al\mbox{.}}{2020}]%
        {aghasadeghi2020zooming}
\bibfield{author}{\bibinfo{person}{Amir Aghasadeghi},
  \bibinfo{person}{Vera~Zaychik Moffitt}, \bibinfo{person}{Sebastian Schelter},
  {and} \bibinfo{person}{Julia Stoyanovich}.} \bibinfo{year}{2020}\natexlab{}.
\newblock \showarticletitle{Zooming Out on an Evolving Graph.}. In
  \bibinfo{booktitle}{\emph{Proc. 23rd Int. Conf. on Extending Database
  Technology}}. \bibinfo{pages}{25--36}.
\newblock


\bibitem[\protect\citeauthoryear{Ammar, McSherry, Salihoglu, and
  Joglekar}{Ammar et~al\mbox{.}}{2018}]%
        {vldb18_AmmarMSJ18}
\bibfield{author}{\bibinfo{person}{Khaled Ammar}, \bibinfo{person}{Frank
  McSherry}, \bibinfo{person}{Semih Salihoglu}, {and} \bibinfo{person}{Manas
  Joglekar}.} \bibinfo{year}{2018}\natexlab{}.
\newblock \showarticletitle{Distributed Evaluation of Subgraph Queries Using
  Worst-case Optimal and Low-Memory Dataflows}.
\newblock \bibinfo{journal}{\emph{Proc. VLDB Endowment}} \bibinfo{volume}{11},
  \bibinfo{number}{6} (\bibinfo{year}{2018}), \bibinfo{pages}{691--704}.
\newblock
\urldef\tempurl%
\url{https://doi.org/10.14778/3184470.3184473}
\showDOI{\tempurl}


\bibitem[\protect\citeauthoryear{Angles, Arenas, Barcelo, Boncz, Fletcher,
  Gutierrez, Lindaaker, Paradies, Plantikow, Sequeda, et~al\mbox{.}}{Angles
  et~al\mbox{.}}{2018}]%
        {angles2018g}
\bibfield{author}{\bibinfo{person}{Renzo Angles}, \bibinfo{person}{Marcelo
  Arenas}, \bibinfo{person}{Pablo Barcelo}, \bibinfo{person}{Peter Boncz},
  \bibinfo{person}{George Fletcher}, \bibinfo{person}{Claudio Gutierrez},
  \bibinfo{person}{Tobias Lindaaker}, \bibinfo{person}{Marcus Paradies},
  \bibinfo{person}{Stefan Plantikow}, \bibinfo{person}{Juan Sequeda},
  {et~al\mbox{.}}} \bibinfo{year}{2018}\natexlab{}.
\newblock \showarticletitle{G-CORE: A core for future graph query languages}.
  In \bibinfo{booktitle}{\emph{Proc. ACM SIGMOD Int. Conf. on Management of
  Data}}. \bibinfo{pages}{1421--1432}.
\newblock


\bibitem[\protect\citeauthoryear{Angles, Arenas, Barcel{\'o}, Hogan, Reutter,
  and Vrgo{\v{c}}}{Angles et~al\mbox{.}}{2017}]%
        {angles2017foundations}
\bibfield{author}{\bibinfo{person}{Renzo Angles}, \bibinfo{person}{Marcelo
  Arenas}, \bibinfo{person}{Pablo Barcel{\'o}}, \bibinfo{person}{Aidan Hogan},
  \bibinfo{person}{Juan Reutter}, {and} \bibinfo{person}{Domagoj Vrgo{\v{c}}}.}
  \bibinfo{year}{2017}\natexlab{}.
\newblock \showarticletitle{Foundations of modern query languages for graph
  databases}.
\newblock \bibinfo{journal}{\emph{ACM Comput. Surv.}} \bibinfo{volume}{50},
  \bibinfo{number}{5} (\bibinfo{year}{2017}), \bibinfo{pages}{68}.
\newblock


\bibitem[\protect\citeauthoryear{Arasu, Babu, and Widom}{Arasu
  et~al\mbox{.}}{2006}]%
        {abw02}
\bibfield{author}{\bibinfo{person}{A. Arasu}, \bibinfo{person}{S. Babu}, {and}
  \bibinfo{person}{J. Widom}.} \bibinfo{year}{2006}\natexlab{}.
\newblock \showarticletitle{The {CQL} Continuous Query Language: Semantic
  Foundations and Query Execution}.
\newblock \bibinfo{journal}{\emph{VLDB J.}} \bibinfo{volume}{15},
  \bibinfo{number}{2} (\bibinfo{year}{2006}), \bibinfo{pages}{121--142}.
\newblock


\bibitem[\protect\citeauthoryear{Babcock, Babu, Datar, Motwani, and
  Widom}{Babcock et~al\mbox{.}}{2002}]%
        {bbd+02}
\bibfield{author}{\bibinfo{person}{B. Babcock}, \bibinfo{person}{S. Babu},
  \bibinfo{person}{M. Datar}, \bibinfo{person}{R. Motwani}, {and}
  \bibinfo{person}{J. Widom}.} \bibinfo{year}{2002}\natexlab{}.
\newblock \showarticletitle{Models and Issues in Data Stream Systems}. In
  \bibinfo{booktitle}{\emph{Proc. ACM SIGACT-SIGMOD Symp. on Principles of
  Database Systems}}. \bibinfo{pages}{1--16}.
\newblock


\bibitem[\protect\citeauthoryear{Baeza}{Baeza}{2013}]%
        {baeza13}
\bibfield{author}{\bibinfo{person}{Pablo~Barcel{\'o} Baeza}.}
  \bibinfo{year}{2013}\natexlab{}.
\newblock \showarticletitle{Querying graph databases}. In
  \bibinfo{booktitle}{\emph{Proc. 32nd ACM SIGACT-SIGMOD-SIGART Symp. on
  Principles of Database Systems}}. \bibinfo{pages}{175--188}.
\newblock


\bibitem[\protect\citeauthoryear{Bagan, Bonifati, Ciucanu, Fletcher, Lemay, and
  Advokaat}{Bagan et~al\mbox{.}}{2016}]%
        {bagan2016gmark}
\bibfield{author}{\bibinfo{person}{Guillaume Bagan}, \bibinfo{person}{Angela
  Bonifati}, \bibinfo{person}{Radu Ciucanu}, \bibinfo{person}{George~HL
  Fletcher}, \bibinfo{person}{Aur{\'e}lien Lemay}, {and} \bibinfo{person}{Nicky
  Advokaat}.} \bibinfo{year}{2016}\natexlab{}.
\newblock \showarticletitle{gMark: schema-driven generation of graphs and
  queries}.
\newblock \bibinfo{journal}{\emph{IEEE Trans. Knowl. and Data Eng.}}
  \bibinfo{volume}{29}, \bibinfo{number}{4} (\bibinfo{year}{2016}),
  \bibinfo{pages}{856--869}.
\newblock


\bibitem[\protect\citeauthoryear{Barbieri, Braga, Ceri, Della~Valle, and
  Grossniklaus}{Barbieri et~al\mbox{.}}{2009}]%
        {barbieri2009c}
\bibfield{author}{\bibinfo{person}{Davide~Francesco Barbieri},
  \bibinfo{person}{Daniele Braga}, \bibinfo{person}{Stefano Ceri},
  \bibinfo{person}{Emanuele Della~Valle}, {and} \bibinfo{person}{Michael
  Grossniklaus}.} \bibinfo{year}{2009}\natexlab{}.
\newblock \showarticletitle{C-SPARQL: SPARQL for continuous querying}. In
  \bibinfo{booktitle}{\emph{Proc. 18th Int. World Wide Web Conf.}}
  \bibinfo{pages}{1061--1062}.
\newblock


\bibitem[\protect\citeauthoryear{Bonifati and Dumbrava}{Bonifati and
  Dumbrava}{2019}]%
        {bonifati2019graph}
\bibfield{author}{\bibinfo{person}{Angela Bonifati} {and}
  \bibinfo{person}{Stefania Dumbrava}.} \bibinfo{year}{2019}\natexlab{}.
\newblock \showarticletitle{Graph queries: From theory to practice}.
\newblock \bibinfo{journal}{\emph{ACM SIGMOD Rec.}} \bibinfo{volume}{47},
  \bibinfo{number}{4} (\bibinfo{year}{2019}), \bibinfo{pages}{5--16}.
\newblock


\bibitem[\protect\citeauthoryear{Bonifati, Fletcher, Voigt, and
  Yakovets}{Bonifati et~al\mbox{.}}{2018}]%
        {bonifati2018querying}
\bibfield{author}{\bibinfo{person}{Angela Bonifati}, \bibinfo{person}{George
  Fletcher}, \bibinfo{person}{Hannes Voigt}, {and} \bibinfo{person}{Nikolay
  Yakovets}.} \bibinfo{year}{2018}\natexlab{}.
\newblock \showarticletitle{Querying Graphs}.
\newblock \bibinfo{journal}{\emph{Synthesis Lectures on Data Management}}
  \bibinfo{volume}{10}, \bibinfo{number}{3} (\bibinfo{year}{2018}),
  \bibinfo{pages}{1--184}.
\newblock


\bibitem[\protect\citeauthoryear{Bonifati, Martens, and Timm}{Bonifati
  et~al\mbox{.}}{2019}]%
        {bonifati2019navigating}
\bibfield{author}{\bibinfo{person}{Angela Bonifati}, \bibinfo{person}{Wim
  Martens}, {and} \bibinfo{person}{Thomas Timm}.}
  \bibinfo{year}{2019}\natexlab{}.
\newblock \showarticletitle{Navigating the Maze of Wikidata Query Logs}. In
  \bibinfo{booktitle}{\emph{Proc. 28th Int. World Wide Web Conf.}}
  \bibinfo{pages}{127--138}.
\newblock


\bibitem[\protect\citeauthoryear{Calbimonte, Corcho, and Gray}{Calbimonte
  et~al\mbox{.}}{2010}]%
        {calbimonte2010enabling}
\bibfield{author}{\bibinfo{person}{Jean-Paul Calbimonte},
  \bibinfo{person}{Oscar Corcho}, {and} \bibinfo{person}{Alasdair~JG Gray}.}
  \bibinfo{year}{2010}\natexlab{}.
\newblock \showarticletitle{Enabling ontology-based access to streaming data
  sources}. In \bibinfo{booktitle}{\emph{Proc. 9th Int. Semantic Web Conf.}}
  \bibinfo{pages}{96--111}.
\newblock


\bibitem[\protect\citeauthoryear{Carbone, Katsifodimos, Ewen, Markl, Haridi,
  and Tzoumas}{Carbone et~al\mbox{.}}{2015}]%
        {carbonekemht15}
\bibfield{author}{\bibinfo{person}{Paris Carbone}, \bibinfo{person}{Asterios
  Katsifodimos}, \bibinfo{person}{Stephan Ewen}, \bibinfo{person}{Volker
  Markl}, \bibinfo{person}{Seif Haridi}, {and} \bibinfo{person}{Kostas
  Tzoumas}.} \bibinfo{year}{2015}\natexlab{}.
\newblock \showarticletitle{Apache Flink{\texttrademark}: Stream and Batch
  Processing in a Single Engine}.
\newblock \bibinfo{journal}{\emph{Q. Bull. IEEE TC on Data Eng.}}
  \bibinfo{volume}{38}, \bibinfo{number}{4} (\bibinfo{year}{2015}),
  \bibinfo{pages}{28--38}.
\newblock
\urldef\tempurl%
\url{http://sites.computer.org/debull/A15dec/p28.pdf}
\showURL{%
\tempurl}


\bibitem[\protect\citeauthoryear{Choudhury, Holder, Jr, Agarwal, and
  Feo}{Choudhury et~al\mbox{.}}{2015}]%
        {edbt15_choudhury:2015aa}
\bibfield{author}{\bibinfo{person}{Sutanay Choudhury},
  \bibinfo{person}{Lawrence~B. Holder}, \bibinfo{person}{George~Chin Jr},
  \bibinfo{person}{Khushbu Agarwal}, {and} \bibinfo{person}{John Feo}.}
  \bibinfo{year}{2015}\natexlab{}.
\newblock \showarticletitle{A {Selectivity} based approach to {Continuous}
  {Pattern} {Detection} in {Streaming} {Graphs}}. In
  \bibinfo{booktitle}{\emph{Proc. 18th Int. Conf. on Extending Database
  Technology}}. \bibinfo{pages}{157--168}.
\newblock
\urldef\tempurl%
\url{https://doi.org/10.5441/002/edbt.2015.15}
\showDOI{\tempurl}


\bibitem[\protect\citeauthoryear{Clifford, Elmasri, Dyreson, Kline, Lorentzos,
  Mitsopoulos, Montanari, Pernici, Sarda, and Segev}{Clifford
  et~al\mbox{.}}{1994}]%
        {clifford1994consensus}
\bibfield{author}{\bibinfo{person}{Christian S Jensen~James Clifford},
  \bibinfo{person}{Ramez Elmasri}, \bibinfo{person}{Curtis Dyreson},
  \bibinfo{person}{Fabio Grandi Wolfgang K\&fer~Nick Kline},
  \bibinfo{person}{Nikos Lorentzos}, \bibinfo{person}{Yamzis Mitsopoulos},
  \bibinfo{person}{Angelo Montanari}, \bibinfo{person}{Daniel Nonen Elisa
  Peressi~Barbara Pernici}, \bibinfo{person}{John F Roddick Nandlal~L Sarda},
  {and} \bibinfo{person}{Maria Rita Scalas~Arie Segev}.}
  \bibinfo{year}{1994}\natexlab{}.
\newblock \showarticletitle{A consensus glossary of temporal database
  concepts}.
\newblock \bibinfo{journal}{\emph{ACM SIGMOD Rec.}} \bibinfo{volume}{23},
  \bibinfo{number}{1} (\bibinfo{year}{1994}).
\newblock


\bibitem[\protect\citeauthoryear{Dell'Aglio, Calbimonte, Della~Valle, and
  Corcho}{Dell'Aglio et~al\mbox{.}}{2015}]%
        {dell2015towards}
\bibfield{author}{\bibinfo{person}{Daniele Dell'Aglio},
  \bibinfo{person}{Jean-Paul Calbimonte}, \bibinfo{person}{Emanuele
  Della~Valle}, {and} \bibinfo{person}{Oscar Corcho}.}
  \bibinfo{year}{2015}\natexlab{}.
\newblock \showarticletitle{Towards a unified language for RDF stream query
  processing}. In \bibinfo{booktitle}{\emph{Proc. 12th Extended Semantic Web
  Conf.}} \bibinfo{pages}{353--363}.
\newblock


\bibitem[\protect\citeauthoryear{Ediger, McColl, Riedy, and Bader}{Ediger
  et~al\mbox{.}}{2012}]%
        {ediger2012stinger}
\bibfield{author}{\bibinfo{person}{David Ediger}, \bibinfo{person}{Rob McColl},
  \bibinfo{person}{Jason Riedy}, {and} \bibinfo{person}{David~A Bader}.}
  \bibinfo{year}{2012}\natexlab{}.
\newblock \showarticletitle{Stinger: High performance data structure for
  streaming graphs}. In \bibinfo{booktitle}{\emph{Proc. 2012 IEEE Conf. on High
  Performance Extreme Comp.}} IEEE, \bibinfo{pages}{1--5}.
\newblock


\bibitem[\protect\citeauthoryear{Erling, Averbuch, Larriba-Pey, Chafi,
  Gubichev, Prat, Pham, and Boncz}{Erling et~al\mbox{.}}{2015}]%
        {sigmod15_erling:2015}
\bibfield{author}{\bibinfo{person}{Orri Erling}, \bibinfo{person}{Alex
  Averbuch}, \bibinfo{person}{Josep Larriba-Pey}, \bibinfo{person}{Hassan
  Chafi}, \bibinfo{person}{Andrey Gubichev}, \bibinfo{person}{Arnau Prat},
  \bibinfo{person}{Minh-Duc Pham}, {and} \bibinfo{person}{Peter Boncz}.}
  \bibinfo{year}{2015}\natexlab{}.
\newblock \showarticletitle{The LDBC Social Network Benchmark: Interactive
  Workload}. In \bibinfo{booktitle}{\emph{Proc. ACM SIGMOD Int. Conf. on
  Management of Data}}. \bibinfo{pages}{619--630}.
\newblock
\showISBNx{978-1-4503-2758-9}
\urldef\tempurl%
\url{https://doi.org/10.1145/2723372.2742786}
\showDOI{\tempurl}


\bibitem[\protect\citeauthoryear{Fan, Hu, and Tian}{Fan et~al\mbox{.}}{2017}]%
        {fan2017incremental}
\bibfield{author}{\bibinfo{person}{Wenfei Fan}, \bibinfo{person}{Chunming Hu},
  {and} \bibinfo{person}{Chao Tian}.} \bibinfo{year}{2017}\natexlab{}.
\newblock \showarticletitle{Incremental graph computations: Doable and
  undoable}. In \bibinfo{booktitle}{\emph{Proc. ACM SIGMOD Int. Conf. on
  Management of Data}}. \bibinfo{pages}{155--169}.
\newblock


\bibitem[\protect\citeauthoryear{Gao, Golab, {\"O}zsu, and Aluc}{Gao
  et~al\mbox{.}}{2018}]%
        {gao:2018aa}
\bibfield{author}{\bibinfo{person}{Libo Gao}, \bibinfo{person}{Lukasz Golab},
  \bibinfo{person}{M.~Tamer {\"O}zsu}, {and} \bibinfo{person}{Gunes Aluc}.}
  \bibinfo{year}{2018}\natexlab{}.
\newblock \showarticletitle{Stream {WatDiv} -- A Streaming {RDF} Benchmark}. In
  \bibinfo{booktitle}{\emph{Proc. ACM SIGMOD Workshop on Semantic Big Data}}.
  \bibinfo{pages}{3:1--3:6}.
\newblock


\bibitem[\protect\citeauthoryear{Ghanem, Hammad, Mokbel, Aref, and
  Elmagarmid}{Ghanem et~al\mbox{.}}{2006}]%
        {ghanem2006incremental}
\bibfield{author}{\bibinfo{person}{Thanaa~M Ghanem},
  \bibinfo{person}{Moustafa~A Hammad}, \bibinfo{person}{Mohamed~F Mokbel},
  \bibinfo{person}{Walid~G Aref}, {and} \bibinfo{person}{Ahmed~K Elmagarmid}.}
  \bibinfo{year}{2006}\natexlab{}.
\newblock \showarticletitle{Incremental evaluation of sliding-window queries
  over data streams}.
\newblock \bibinfo{journal}{\emph{IEEE Trans. Knowl. and Data Eng.}}
  \bibinfo{volume}{19}, \bibinfo{number}{1} (\bibinfo{year}{2006}),
  \bibinfo{pages}{57--72}.
\newblock


\bibitem[\protect\citeauthoryear{Golab}{Golab}{2006}]%
        {golab:2006yq}
\bibfield{author}{\bibinfo{person}{L. Golab}.} \bibinfo{year}{2006}\natexlab{}.
\newblock \emph{\bibinfo{title}{Sliding Window Query Processing over Data
  Streams}}.
\newblock \bibinfo{thesistype}{Ph.D. Dissertation}. \bibinfo{school}{University
  of Waterloo}.
\newblock


\bibitem[\protect\citeauthoryear{Golab and {\"O}zsu}{Golab and
  {\"O}zsu}{2003a}]%
        {golabo03a}
\bibfield{author}{\bibinfo{person}{Lukasz Golab} {and}
  \bibinfo{person}{M.~Tamer {\"O}zsu}.} \bibinfo{year}{2003}\natexlab{a}.
\newblock \showarticletitle{Issues in data stream management}.
\newblock \bibinfo{journal}{\emph{ACM SIGMOD Rec.}} \bibinfo{volume}{32},
  \bibinfo{number}{2} (\bibinfo{year}{2003}), \bibinfo{pages}{5--14}.
\newblock


\bibitem[\protect\citeauthoryear{Golab and {\"O}zsu}{Golab and
  {\"O}zsu}{2003b}]%
        {vldb03_golabo03}
\bibfield{author}{\bibinfo{person}{Lukasz Golab} {and}
  \bibinfo{person}{M.~Tamer {\"O}zsu}.} \bibinfo{year}{2003}\natexlab{b}.
\newblock \showarticletitle{Processing Sliding Window Multi-Joins in Continuous
  Queries over Data Streams}. In \bibinfo{booktitle}{\emph{Proc. 29th Int.
  Conf. on Very Large Data Bases}}. \bibinfo{pages}{500--511}.
\newblock


\bibitem[\protect\citeauthoryear{Golab and {\"O}zsu}{Golab and
  {\"O}zsu}{2005}]%
        {golabo05}
\bibfield{author}{\bibinfo{person}{Lukasz Golab} {and}
  \bibinfo{person}{M.~Tamer {\"O}zsu}.} \bibinfo{year}{2005}\natexlab{}.
\newblock \showarticletitle{Update-Pattern-Aware Modeling and Processing of
  Continuous Queries}. In \bibinfo{booktitle}{\emph{Proc. ACM SIGMOD Int. Conf.
  on Management of Data}}. \bibinfo{pages}{658--669}.
\newblock


\bibitem[\protect\citeauthoryear{Graefe}{Graefe}{1993}]%
        {graefe93}
\bibfield{author}{\bibinfo{person}{G. Graefe}.}
  \bibinfo{year}{1993}\natexlab{}.
\newblock \showarticletitle{Query Evaluation Techniques for Large Databases}.
\newblock \bibinfo{journal}{\emph{ACM Comput. Surv.}} \bibinfo{volume}{25},
  \bibinfo{number}{2} (\bibinfo{year}{1993}), \bibinfo{pages}{73--170}.
\newblock


\bibitem[\protect\citeauthoryear{Grewal, Jiang, Lam, Jung, Vuddemarri, Li,
  Landge, and Lin}{Grewal et~al\mbox{.}}{2018}]%
        {grewal2018recservice}
\bibfield{author}{\bibinfo{person}{Ajeet Grewal}, \bibinfo{person}{Jerry
  Jiang}, \bibinfo{person}{Gary Lam}, \bibinfo{person}{Tristan Jung},
  \bibinfo{person}{Lohith Vuddemarri}, \bibinfo{person}{Quannan Li},
  \bibinfo{person}{Aaditya Landge}, {and} \bibinfo{person}{Jimmy Lin}.}
  \bibinfo{year}{2018}\natexlab{}.
\newblock \showarticletitle{RecService: Multi-Tenant Distributed Real-Time
  Graph Processing at Twitter}. In \bibinfo{booktitle}{\emph{Proc. 10th USENIX
  Workshop on Hot Topics in Cloud Computing}}.
\newblock


\bibitem[\protect\citeauthoryear{Gupta, Mumick, and Subrahmanian}{Gupta
  et~al\mbox{.}}{1993}]%
        {gms93}
\bibfield{author}{\bibinfo{person}{A. Gupta}, \bibinfo{person}{I.~S. Mumick},
  {and} \bibinfo{person}{V.~S. Subrahmanian}.} \bibinfo{year}{1993}\natexlab{}.
\newblock \showarticletitle{Maintaining Views Incrementally}. In
  \bibinfo{booktitle}{\emph{Proc. ACM SIGMOD Int. Conf. on Management of
  Data}}. \bibinfo{pages}{157--166}.
\newblock


\bibitem[\protect\citeauthoryear{Hirzel, Baudart, Bonifati, Valle, Sakr, and
  Vlachou}{Hirzel et~al\mbox{.}}{2018}]%
        {HirzelBBVSV18}
\bibfield{author}{\bibinfo{person}{Martin Hirzel}, \bibinfo{person}{Guillaume
  Baudart}, \bibinfo{person}{Angela Bonifati}, \bibinfo{person}{Emanuele~Della
  Valle}, \bibinfo{person}{Sherif Sakr}, {and} \bibinfo{person}{Akrivi
  Vlachou}.} \bibinfo{year}{2018}\natexlab{}.
\newblock \showarticletitle{Stream Processing Languages in the Big Data Era}.
\newblock \bibinfo{journal}{\emph{ACM SIGMOD Rec.}} \bibinfo{volume}{47},
  \bibinfo{number}{2} (\bibinfo{year}{2018}), \bibinfo{pages}{29--40}.
\newblock


\bibitem[\protect\citeauthoryear{Iyer, Li, Das, and Stoica}{Iyer
  et~al\mbox{.}}{2016}]%
        {iyer2016time}
\bibfield{author}{\bibinfo{person}{Anand~Padmanabha Iyer},
  \bibinfo{person}{Li~Erran Li}, \bibinfo{person}{Tathagata Das}, {and}
  \bibinfo{person}{Ion Stoica}.} \bibinfo{year}{2016}\natexlab{}.
\newblock \showarticletitle{Time-evolving graph processing at scale}. In
  \bibinfo{booktitle}{\emph{Proc. 4th Int. Workshop on Graph Data Management
  Experiences and Systems}}. \bibinfo{pages}{1--6}.
\newblock


\bibitem[\protect\citeauthoryear{Kim, Seo, Han, Lee, Hong, Chafi, Shin, and
  Jeong}{Kim et~al\mbox{.}}{2018}]%
        {kim2018turboflux}
\bibfield{author}{\bibinfo{person}{Kyoungmin Kim}, \bibinfo{person}{In Seo},
  \bibinfo{person}{Wook-Shin Han}, \bibinfo{person}{Jeong-Hoon Lee},
  \bibinfo{person}{Sungpack Hong}, \bibinfo{person}{Hassan Chafi},
  \bibinfo{person}{Hyungyu Shin}, {and} \bibinfo{person}{Geonhwa Jeong}.}
  \bibinfo{year}{2018}\natexlab{}.
\newblock \showarticletitle{TurboFlux: A Fast Continuous Subgraph Matching
  System for Streaming Graph Data}. In \bibinfo{booktitle}{\emph{Proc. ACM
  SIGMOD Int. Conf. on Management of Data}}. \bibinfo{pages}{411--426}.
\newblock


\bibitem[\protect\citeauthoryear{Koch, Ahmad, Kennedy, Nikolic, N{\"{o}}tzli,
  Lupei, and Shaikhha}{Koch et~al\mbox{.}}{2014}]%
        {KochAKNNLS14}
\bibfield{author}{\bibinfo{person}{Christoph Koch}, \bibinfo{person}{Yanif
  Ahmad}, \bibinfo{person}{Oliver Kennedy}, \bibinfo{person}{Milos Nikolic},
  \bibinfo{person}{Andres N{\"{o}}tzli}, \bibinfo{person}{Daniel Lupei}, {and}
  \bibinfo{person}{Amir Shaikhha}.} \bibinfo{year}{2014}\natexlab{}.
\newblock \showarticletitle{{DBToaster:} Higher-Order Delta Processing for
  Dynamic, Frequently Fresh Views}.
\newblock \bibinfo{journal}{\emph{VLDB J.}} \bibinfo{volume}{23},
  \bibinfo{number}{2} (\bibinfo{year}{2014}), \bibinfo{pages}{253--278}.
\newblock


\bibitem[\protect\citeauthoryear{Komazec, Cerri, and Fensel}{Komazec
  et~al\mbox{.}}{2012}]%
        {komazec2012sparkwave}
\bibfield{author}{\bibinfo{person}{Srdjan Komazec}, \bibinfo{person}{Davide
  Cerri}, {and} \bibinfo{person}{Dieter Fensel}.}
  \bibinfo{year}{2012}\natexlab{}.
\newblock \showarticletitle{Sparkwave: continuous schema-enhanced pattern
  matching over RDF data streams}. In \bibinfo{booktitle}{\emph{Proc. 6th Int.
  Conf. Distributed Event-Based Systems}}. \bibinfo{pages}{58--68}.
\newblock


\bibitem[\protect\citeauthoryear{Koschmieder and Leser}{Koschmieder and
  Leser}{2012}]%
        {koschmieder2012regular}
\bibfield{author}{\bibinfo{person}{Andr{\'e} Koschmieder} {and}
  \bibinfo{person}{Ulf Leser}.} \bibinfo{year}{2012}\natexlab{}.
\newblock \showarticletitle{Regular path queries on large graphs}. In
  \bibinfo{booktitle}{\emph{SSDBM12}}. \bibinfo{pages}{177--194}.
\newblock


\bibitem[\protect\citeauthoryear{Kr{\"a}mer and Seeger}{Kr{\"a}mer and
  Seeger}{2009}]%
        {kramer2009semantics}
\bibfield{author}{\bibinfo{person}{J{\"u}rgen Kr{\"a}mer} {and}
  \bibinfo{person}{Bernhard Seeger}.} \bibinfo{year}{2009}\natexlab{}.
\newblock \showarticletitle{Semantics and implementation of continuous sliding
  window queries over data streams}.
\newblock \bibinfo{journal}{\emph{ACM Trans. Database Syst.}}
  \bibinfo{volume}{34}, \bibinfo{number}{1} (\bibinfo{year}{2009}),
  \bibinfo{pages}{1--49}.
\newblock


\bibitem[\protect\citeauthoryear{Kulkarni, Bhagat, Fu, Kedigehalli, Kellogg,
  Mittal, Patel, Ramasamy, and Taneja}{Kulkarni et~al\mbox{.}}{2015}]%
        {sigmod15_kulkarni:2015}
\bibfield{author}{\bibinfo{person}{Sanjeev Kulkarni}, \bibinfo{person}{Nikunj
  Bhagat}, \bibinfo{person}{Maosong Fu}, \bibinfo{person}{Vikas Kedigehalli},
  \bibinfo{person}{Christopher Kellogg}, \bibinfo{person}{Sailesh Mittal},
  \bibinfo{person}{Jignesh~M. Patel}, \bibinfo{person}{Karthik Ramasamy}, {and}
  \bibinfo{person}{Siddarth Taneja}.} \bibinfo{year}{2015}\natexlab{}.
\newblock \showarticletitle{Twitter Heron: Stream Processing at Scale}. In
  \bibinfo{booktitle}{\emph{Proc. ACM SIGMOD Int. Conf. on Management of
  Data}}. \bibinfo{pages}{239--250}.
\newblock
\showISBNx{978-1-4503-2758-9}
\urldef\tempurl%
\url{https://doi.org/10.1145/2723372.2742788}
\showDOI{\tempurl}


\bibitem[\protect\citeauthoryear{Kumar and Huang}{Kumar and Huang}{2019}]%
        {kumar2019graphone}
\bibfield{author}{\bibinfo{person}{Pradeep Kumar} {and}
  \bibinfo{person}{H~Howie Huang}.} \bibinfo{year}{2019}\natexlab{}.
\newblock \showarticletitle{GraphOne: A data store for real-time analytics on
  evolving graphs}. In \bibinfo{booktitle}{\emph{Proc. 17th USENIX Conf. on
  File and Storage Technologies}}. \bibinfo{pages}{249--263}.
\newblock


\bibitem[\protect\citeauthoryear{Kumar and Huang}{Kumar and Huang}{2020}]%
        {kumar2020graphone}
\bibfield{author}{\bibinfo{person}{Pradeep Kumar} {and}
  \bibinfo{person}{H~Howie Huang}.} \bibinfo{year}{2020}\natexlab{}.
\newblock \showarticletitle{GraphOne: A Data Store for Real-time Analytics on
  Evolving Graphs}.
\newblock \bibinfo{journal}{\emph{ACM Trans. Storage}} \bibinfo{volume}{15},
  \bibinfo{number}{4} (\bibinfo{year}{2020}), \bibinfo{pages}{1--40}.
\newblock


\bibitem[\protect\citeauthoryear{Le-Phuoc, Dao-Tran, Parreira, and
  Hauswirth}{Le-Phuoc et~al\mbox{.}}{2011}]%
        {le2011native}
\bibfield{author}{\bibinfo{person}{Danh Le-Phuoc}, \bibinfo{person}{Minh
  Dao-Tran}, \bibinfo{person}{Josiane~Xavier Parreira}, {and}
  \bibinfo{person}{Manfred Hauswirth}.} \bibinfo{year}{2011}\natexlab{}.
\newblock \showarticletitle{A native and adaptive approach for unified
  processing of linked streams and linked data}. In
  \bibinfo{booktitle}{\emph{Proc. 10th Int. Semantic Web Conf.}}
  \bibinfo{pages}{370--388}.
\newblock


\bibitem[\protect\citeauthoryear{Li, Zou, {\"O}zsu, and Zhao}{Li
  et~al\mbox{.}}{2019}]%
        {li2019time}
\bibfield{author}{\bibinfo{person}{Youhuan Li}, \bibinfo{person}{Lei Zou},
  \bibinfo{person}{M~Tamer {\"O}zsu}, {and} \bibinfo{person}{Dongyan Zhao}.}
  \bibinfo{year}{2019}\natexlab{}.
\newblock \showarticletitle{Time constrained continuous subgraph search over
  streaming graphs}. In \bibinfo{booktitle}{\emph{Proc. 35th Int. Conf. on Data
  Engineering}}. IEEE Press, \bibinfo{pages}{1082--1093}.
\newblock


\bibitem[\protect\citeauthoryear{Libkin, Reutter, Soto, and Vrgo{\v{c}}}{Libkin
  et~al\mbox{.}}{2018}]%
        {libkin2018trial}
\bibfield{author}{\bibinfo{person}{Leonid Libkin}, \bibinfo{person}{Juan~L
  Reutter}, \bibinfo{person}{Adrian Soto}, {and} \bibinfo{person}{Domagoj
  Vrgo{\v{c}}}.} \bibinfo{year}{2018}\natexlab{}.
\newblock \showarticletitle{TriAL: A navigational algebra for RDF
  triplestores}.
\newblock \bibinfo{journal}{\emph{ACM Trans. Database Syst.}}
  \bibinfo{volume}{43}, \bibinfo{number}{1} (\bibinfo{year}{2018}),
  \bibinfo{pages}{1--46}.
\newblock


\bibitem[\protect\citeauthoryear{Libkin and Vrgo{\v{c}}}{Libkin and
  Vrgo{\v{c}}}{2012}]%
        {libkin2012regular}
\bibfield{author}{\bibinfo{person}{Leonid Libkin} {and}
  \bibinfo{person}{Domagoj Vrgo{\v{c}}}.} \bibinfo{year}{2012}\natexlab{}.
\newblock \showarticletitle{Regular path queries on graphs with data}. In
  \bibinfo{booktitle}{\emph{Proc. 15th Int. Conf. on Database Theory}}.
  \bibinfo{pages}{74--85}.
\newblock


\bibitem[\protect\citeauthoryear{Liu and {\"O}zsu}{Liu and {\"O}zsu}{2009}]%
        {lingozsu09}
\bibfield{editor}{\bibinfo{person}{Ling Liu} {and} \bibinfo{person}{M.~Tamer
  {\"O}zsu}} (Eds.). \bibinfo{year}{2009}\natexlab{}.
\newblock \bibinfo{booktitle}{\emph{Encyclopedia of Database Systems}}.
\newblock \bibinfo{publisher}{{Springer}}.
\newblock


\bibitem[\protect\citeauthoryear{Liu, Taylor, Zhou, Ives, and Loo}{Liu
  et~al\mbox{.}}{2009}]%
        {icde09_4812481}
\bibfield{author}{\bibinfo{person}{Mengmeng Liu}, \bibinfo{person}{N.E.
  Taylor}, \bibinfo{person}{Wenchao Zhou}, \bibinfo{person}{Z.G. Ives}, {and}
  \bibinfo{person}{Boon~Thau Loo}.} \bibinfo{year}{2009}\natexlab{}.
\newblock \showarticletitle{Recursive Computation of Regions and Connectivity
  in Networks}. In \bibinfo{booktitle}{\emph{Proc. 25th Int. Conf. on Data
  Engineering}}. \bibinfo{pages}{1108--1119}.
\newblock
\showISSN{1084-4627}
\urldef\tempurl%
\url{https://doi.org/10.1109/ICDE.2009.36}
\showDOI{\tempurl}


\bibitem[\protect\citeauthoryear{Mariappan and Vora}{Mariappan and
  Vora}{2019}]%
        {mariappan2019graphbolt}
\bibfield{author}{\bibinfo{person}{Mugilan Mariappan} {and}
  \bibinfo{person}{Keval Vora}.} \bibinfo{year}{2019}\natexlab{}.
\newblock \showarticletitle{GraphBolt: Dependency-driven synchronous processing
  of streaming graphs}. In \bibinfo{booktitle}{\emph{Proc. 14th ACM
  SIGOPS/EuroSys European Conf. on Comp. Syst.}} \bibinfo{pages}{1--16}.
\newblock


\bibitem[\protect\citeauthoryear{McSherry, Lattuada, Schwarzkopf, and
  Roscoe}{McSherry et~al\mbox{.}}{2020}]%
        {mcherry_shared_vldb20}
\bibfield{author}{\bibinfo{person}{Frank McSherry}, \bibinfo{person}{Andrea
  Lattuada}, \bibinfo{person}{Malte Schwarzkopf}, {and}
  \bibinfo{person}{Timothy Roscoe}.} \bibinfo{year}{2020}\natexlab{}.
\newblock \showarticletitle{Shared Arrangements: Practical Inter-Query Sharing
  for Streaming Dataflows}.
\newblock \bibinfo{journal}{\emph{Proc. VLDB Endowment}} \bibinfo{volume}{13},
  \bibinfo{number}{10} (\bibinfo{year}{2020}), \bibinfo{pages}{1793--1806}.
\newblock


\bibitem[\protect\citeauthoryear{McSherry, Murray, Isaacs, and Isard}{McSherry
  et~al\mbox{.}}{2013}]%
        {cidr13_mcsherrymii13}
\bibfield{author}{\bibinfo{person}{Frank McSherry},
  \bibinfo{person}{Derek~Gordon Murray}, \bibinfo{person}{Rebecca Isaacs},
  {and} \bibinfo{person}{Michael Isard}.} \bibinfo{year}{2013}\natexlab{}.
\newblock \showarticletitle{Differential Dataflow}. In
  \bibinfo{booktitle}{\emph{Proc. 6th Biennial Conf. on Innovative Data Systems
  Research}}.
\newblock


\bibitem[\protect\citeauthoryear{Mendelzon and Wood}{Mendelzon and
  Wood}{1995}]%
        {mendelzon1995finding}
\bibfield{author}{\bibinfo{person}{Alberto~O Mendelzon} {and}
  \bibinfo{person}{Peter~T Wood}.} \bibinfo{year}{1995}\natexlab{}.
\newblock \showarticletitle{Finding regular simple paths in graph databases}.
\newblock \bibinfo{journal}{\emph{SIAM J. on Comput.}} \bibinfo{volume}{24},
  \bibinfo{number}{6} (\bibinfo{year}{1995}), \bibinfo{pages}{1235--1258}.
\newblock


\bibitem[\protect\citeauthoryear{Moffitt and Stoyanovich}{Moffitt and
  Stoyanovich}{2017}]%
        {moffitt2017temporal}
\bibfield{author}{\bibinfo{person}{Vera~Zaychik Moffitt} {and}
  \bibinfo{person}{Julia Stoyanovich}.} \bibinfo{year}{2017}\natexlab{}.
\newblock \showarticletitle{Temporal graph algebra}. In
  \bibinfo{booktitle}{\emph{Proc. 16th Int. Symposium on Database Programming
  Languages}}. \bibinfo{pages}{1--12}.
\newblock


\bibitem[\protect\citeauthoryear{Murray, McSherry, Isaacs, Isard, Barham, and
  Abadi}{Murray et~al\mbox{.}}{2013}]%
        {murray2013naiad}
\bibfield{author}{\bibinfo{person}{Derek~G Murray}, \bibinfo{person}{Frank
  McSherry}, \bibinfo{person}{Rebecca Isaacs}, \bibinfo{person}{Michael Isard},
  \bibinfo{person}{Paul Barham}, {and} \bibinfo{person}{Mart{\'\i}n Abadi}.}
  \bibinfo{year}{2013}\natexlab{}.
\newblock \showarticletitle{Naiad: a timely dataflow system}. In
  \bibinfo{booktitle}{\emph{Proc. 24th ACM Symp. on Operating System
  Principles}}. \bibinfo{pages}{439--455}.
\newblock


\bibitem[\protect\citeauthoryear{Ngo, R{\'e}, and Rudra}{Ngo
  et~al\mbox{.}}{2014}]%
        {Ngo_2014}
\bibfield{author}{\bibinfo{person}{Hung~Q Ngo}, \bibinfo{person}{Christopher
  R{\'e}}, {and} \bibinfo{person}{Atri Rudra}.}
  \bibinfo{year}{2014}\natexlab{}.
\newblock \showarticletitle{Skew strikes back}.
\newblock \bibinfo{journal}{\emph{ACM SIGMOD Rec.}} \bibinfo{volume}{42},
  \bibinfo{number}{4} (\bibinfo{date}{Feb} \bibinfo{year}{2014}),
  \bibinfo{pages}{5--16}.
\newblock
\showISSN{0163-5808}
\urldef\tempurl%
\url{https://doi.org/10.1145/2590989.2590991}
\showDOI{\tempurl}


\bibitem[\protect\citeauthoryear{Nikolic and Olteanu}{Nikolic and
  Olteanu}{2018}]%
        {nikolic2018incremental}
\bibfield{author}{\bibinfo{person}{Milos Nikolic} {and} \bibinfo{person}{Dan
  Olteanu}.} \bibinfo{year}{2018}\natexlab{}.
\newblock \showarticletitle{Incremental view maintenance with triple lock
  factorization benefits}. In \bibinfo{booktitle}{\emph{Proc. ACM SIGMOD Int.
  Conf. on Management of Data}}. \bibinfo{pages}{365--380}.
\newblock


\bibitem[\protect\citeauthoryear{Pacaci, Bonifati, and {\"O}zsu}{Pacaci
  et~al\mbox{.}}{2020}]%
        {Pacaci_2020}
\bibfield{author}{\bibinfo{person}{Anil Pacaci}, \bibinfo{person}{Angela
  Bonifati}, {and} \bibinfo{person}{M.~Tamer {\"O}zsu}.}
  \bibinfo{year}{2020}\natexlab{}.
\newblock \showarticletitle{Regular Path Query Evaluation on Streaming Graphs}.
  In \bibinfo{booktitle}{\emph{Proc. ACM SIGMOD Int. Conf. on Management of
  Data}}. ACM, \bibinfo{pages}{1415--1430}.
\newblock
\showISBNx{9781450367356}
\urldef\tempurl%
\url{https://doi.org/10.1145/3318464.3389733}
\showDOI{\tempurl}


\bibitem[\protect\citeauthoryear{Pacaci, Zhou, Lin, and {\"O}zsu}{Pacaci
  et~al\mbox{.}}{2017}]%
        {pacaci:2017aa}
\bibfield{author}{\bibinfo{person}{Anil Pacaci}, \bibinfo{person}{Alice Zhou},
  \bibinfo{person}{Jimmy Lin}, {and} \bibinfo{person}{M.~Tamer {\"O}zsu}.}
  \bibinfo{year}{2017}\natexlab{}.
\newblock \showarticletitle{Do We Need Specialized Graph Databases?:
  Benchmarking Real-Time Social Networking Applications}. In
  \bibinfo{booktitle}{\emph{Proc. 5th Int. Workshop on Graph Data Management
  Experiences and Systems}} (Chicago, IL, USA). Article
  \bibinfo{articleno}{12}, \bibinfo{numpages}{7}~pages.
\newblock
\showISBNx{978-1-4503-5038-9}
\urldef\tempurl%
\url{https://doi.org/10.1145/3078447.3078459}
\showDOI{\tempurl}


\bibitem[\protect\citeauthoryear{Paranjape, Benson, and Leskovec}{Paranjape
  et~al\mbox{.}}{2017}]%
        {paranjape2017motifs}
\bibfield{author}{\bibinfo{person}{Ashwin Paranjape}, \bibinfo{person}{Austin~R
  Benson}, {and} \bibinfo{person}{Jure Leskovec}.}
  \bibinfo{year}{2017}\natexlab{}.
\newblock \showarticletitle{Motifs in temporal networks}. In
  \bibinfo{booktitle}{\emph{Proc. 10th ACM Int. Conf. Web Search and Data
  Mining}}. \bibinfo{pages}{601--610}.
\newblock


\bibitem[\protect\citeauthoryear{Qiu, Cen, Qian, Peng, Zhang, Lin, and
  Zhou}{Qiu et~al\mbox{.}}{2018}]%
        {qiu2018real}
\bibfield{author}{\bibinfo{person}{Xiafei Qiu}, \bibinfo{person}{Wubin Cen},
  \bibinfo{person}{Zhengping Qian}, \bibinfo{person}{You Peng},
  \bibinfo{person}{Ying Zhang}, \bibinfo{person}{Xuemin Lin}, {and}
  \bibinfo{person}{Jingren Zhou}.} \bibinfo{year}{2018}\natexlab{}.
\newblock \showarticletitle{Real-time constrained cycle detection in large
  dynamic graphs}.
\newblock \bibinfo{journal}{\emph{Proc. VLDB Endowment}} \bibinfo{volume}{11},
  \bibinfo{number}{12} (\bibinfo{year}{2018}), \bibinfo{pages}{1876--1888}.
\newblock


\bibitem[\protect\citeauthoryear{Reutter, Romero, and Vardi}{Reutter
  et~al\mbox{.}}{2017}]%
        {reutter2017regular}
\bibfield{author}{\bibinfo{person}{Juan~L Reutter}, \bibinfo{person}{Miguel
  Romero}, {and} \bibinfo{person}{Moshe~Y Vardi}.}
  \bibinfo{year}{2017}\natexlab{}.
\newblock \showarticletitle{Regular queries on graph databases}.
\newblock \bibinfo{journal}{\emph{Theory of Comput. Syst.}}
  \bibinfo{volume}{61}, \bibinfo{number}{1} (\bibinfo{year}{2017}),
  \bibinfo{pages}{31--83}.
\newblock


\bibitem[\protect\citeauthoryear{Sahu, Mhedhbi, Salihoglu, Lin, and
  {\"{O}}zsu}{Sahu et~al\mbox{.}}{2020}]%
        {vldbj19_Sahu:2017aa}
\bibfield{author}{\bibinfo{person}{Siddhartha Sahu}, \bibinfo{person}{Amine
  Mhedhbi}, \bibinfo{person}{Semih Salihoglu}, \bibinfo{person}{Jimmy Lin},
  {and} \bibinfo{person}{M.~Tamer {\"{O}}zsu}.}
  \bibinfo{year}{2020}\natexlab{}.
\newblock \showarticletitle{The Ubiquity of Large Graphs and Surprising
  Challenges of Graph Processing}.
\newblock \bibinfo{journal}{\emph{VLDB J.}}  \bibinfo{volume}{29}
  (\bibinfo{year}{2020}), \bibinfo{pages}{595---618}.
\newblock
\urldef\tempurl%
\url{https://doi.org/10.1007/s00778-019-00548-x}
\showDOI{\tempurl}


\bibitem[\protect\citeauthoryear{Sahu and Salihoglu}{Sahu and
  Salihoglu}{2020}]%
        {graphsurge2020}
\bibfield{author}{\bibinfo{person}{Siddhartha Sahu} {and}
  \bibinfo{person}{Semih Salihoglu}.} \bibinfo{year}{2020}\natexlab{}.
\newblock \showarticletitle{Graphsurge: Graph Analytics on View Collections
  Using Differential Computation}.
\newblock \bibinfo{journal}{\emph{arXiv preprint arXiv:2004.05297}}
  (\bibinfo{year}{2020}).
\newblock


\bibitem[\protect\citeauthoryear{Sakr, Bonifati, Voigt, Iosup, Ammar, Angles,
  Aref, Arenas, Besta, Boncz, Daudjee, Valle, Dumbrava, Hartig, Haslhofer,
  Hegeman, Hidders, Hose, Iamnitchi, Kalavri, Kapp, Martens, {\"{O}}zsu,
  Peukert, Plantikow, Ragab, Ripeanu, Salihoglu, Schulz, Selmer, Sequeda,
  Shinavier, Sz{\'{a}}rnyas, Tommasini, Tumeo, Uta, Varbanescu, Wu, Yakovets,
  Yan, and Yoneki}{Sakr et~al\mbox{.}}{2020}]%
        {abs-2012-06171}
\bibfield{author}{\bibinfo{person}{Sherif Sakr}, \bibinfo{person}{Angela
  Bonifati}, \bibinfo{person}{Hannes Voigt}, \bibinfo{person}{Alexandru Iosup},
  \bibinfo{person}{Khaled Ammar}, \bibinfo{person}{Renzo Angles},
  \bibinfo{person}{Walid~G. Aref}, \bibinfo{person}{Marcelo Arenas},
  \bibinfo{person}{Maciej Besta}, \bibinfo{person}{Peter~A. Boncz},
  \bibinfo{person}{Khuzaima Daudjee}, \bibinfo{person}{Emanuele~Della Valle},
  \bibinfo{person}{Stefania Dumbrava}, \bibinfo{person}{Olaf Hartig},
  \bibinfo{person}{Bernhard Haslhofer}, \bibinfo{person}{Tim Hegeman},
  \bibinfo{person}{Jan Hidders}, \bibinfo{person}{Katja Hose},
  \bibinfo{person}{Adriana Iamnitchi}, \bibinfo{person}{Vasiliki Kalavri},
  \bibinfo{person}{Hugo Kapp}, \bibinfo{person}{Wim Martens},
  \bibinfo{person}{M.~Tamer {\"{O}}zsu}, \bibinfo{person}{Eric Peukert},
  \bibinfo{person}{Stefan Plantikow}, \bibinfo{person}{Mohamed Ragab},
  \bibinfo{person}{Matei Ripeanu}, \bibinfo{person}{Semih Salihoglu},
  \bibinfo{person}{Christian Schulz}, \bibinfo{person}{Petra Selmer},
  \bibinfo{person}{Juan~F. Sequeda}, \bibinfo{person}{Joshua Shinavier},
  \bibinfo{person}{G{\'{a}}bor Sz{\'{a}}rnyas}, \bibinfo{person}{Riccardo
  Tommasini}, \bibinfo{person}{Antonino Tumeo}, \bibinfo{person}{Alexandru
  Uta}, \bibinfo{person}{Ana~Lucia Varbanescu}, \bibinfo{person}{Hsiang{-}Yun
  Wu}, \bibinfo{person}{Nikolay Yakovets}, \bibinfo{person}{Da Yan}, {and}
  \bibinfo{person}{Eiko Yoneki}.} \bibinfo{year}{2020}\natexlab{}.
\newblock \showarticletitle{The Future is Big Graphs! {A} Community View on
  Graph Processing Systems}.
\newblock \bibinfo{journal}{\emph{CoRR}}  \bibinfo{volume}{abs/2012.06171}
  (\bibinfo{year}{2020}).
\newblock


\bibitem[\protect\citeauthoryear{Salihoglu and Yakovets}{Salihoglu and
  Yakovets}{2019}]%
        {salihoglu2019graph}
\bibfield{author}{\bibinfo{person}{Semih Salihoglu} {and}
  \bibinfo{person}{Nikolay Yakovets}.} \bibinfo{year}{2019}\natexlab{}.
\newblock \showarticletitle{Graph query processing}.
\newblock \bibinfo{journal}{\emph{Encyclopedia of Big Data Technologies}}
  (\bibinfo{year}{2019}), \bibinfo{pages}{890--898}.
\newblock


\bibitem[\protect\citeauthoryear{Sengupta, Sundaram, Zhu, Willke, Young, Wolf,
  and Schwan}{Sengupta et~al\mbox{.}}{2016}]%
        {sengupta2016graphin}
\bibfield{author}{\bibinfo{person}{Dipanjan Sengupta},
  \bibinfo{person}{Narayanan Sundaram}, \bibinfo{person}{Xia Zhu},
  \bibinfo{person}{Theodore~L Willke}, \bibinfo{person}{Jeffrey Young},
  \bibinfo{person}{Matthew Wolf}, {and} \bibinfo{person}{Karsten Schwan}.}
  \bibinfo{year}{2016}\natexlab{}.
\newblock \showarticletitle{Graphin: An online high performance incremental
  graph processing framework}. In \bibinfo{booktitle}{\emph{European Conference
  on Parallel Processing}}. Springer, \bibinfo{pages}{319--333}.
\newblock


\bibitem[\protect\citeauthoryear{Sheng, Cao, Cai, Yao, and Xie}{Sheng
  et~al\mbox{.}}{2018}]%
        {sheng2018grapu}
\bibfield{author}{\bibinfo{person}{Feng Sheng}, \bibinfo{person}{Qiang Cao},
  \bibinfo{person}{Haoran Cai}, \bibinfo{person}{Jie Yao}, {and}
  \bibinfo{person}{Changsheng Xie}.} \bibinfo{year}{2018}\natexlab{}.
\newblock \showarticletitle{GraPU: Accelerate streaming graph analysis through
  preprocessing buffered updates}. In \bibinfo{booktitle}{\emph{Proc. 9th ACM
  Symp. on Cloud Computing}}. \bibinfo{pages}{301--312}.
\newblock


\bibitem[\protect\citeauthoryear{Toshniwal, Taneja, Shukla, Ramasamy, Patel,
  Kulkarni, Jackson, Gade, Fu, Donham, et~al\mbox{.}}{Toshniwal
  et~al\mbox{.}}{2014}]%
        {toshniwal2014storm}
\bibfield{author}{\bibinfo{person}{Ankit Toshniwal}, \bibinfo{person}{Siddarth
  Taneja}, \bibinfo{person}{Amit Shukla}, \bibinfo{person}{Karthik Ramasamy},
  \bibinfo{person}{Jignesh~M Patel}, \bibinfo{person}{Sanjeev Kulkarni},
  \bibinfo{person}{Jason Jackson}, \bibinfo{person}{Krishna Gade},
  \bibinfo{person}{Maosong Fu}, \bibinfo{person}{Jake Donham}, {et~al\mbox{.}}}
  \bibinfo{year}{2014}\natexlab{}.
\newblock \showarticletitle{Storm@ twitter}. In \bibinfo{booktitle}{\emph{Proc.
  ACM SIGMOD Int. Conf. on Management of Data}}. \bibinfo{pages}{147--156}.
\newblock


\bibitem[\protect\citeauthoryear{Urhan and Franklin}{Urhan and
  Franklin}{2000}]%
        {urhanf2000}
\bibfield{author}{\bibinfo{person}{Tolga Urhan} {and} \bibinfo{person}{M.J.
  Franklin}.} \bibinfo{year}{2000}\natexlab{}.
\newblock \showarticletitle{{XJoin}: A Reactively-Scheduled Pipelined Join
  Operator}.
\newblock \bibinfo{journal}{\emph{Q. Bull. IEEE TC on Data Eng.}}
  \bibinfo{volume}{23} (\bibinfo{year}{2000}), \bibinfo{pages}{27}.
\newblock


\bibitem[\protect\citeauthoryear{van Rest, Hong, Kim, Meng, and Chafi}{van Rest
  et~al\mbox{.}}{2016}]%
        {grades2016_van2016pgql}
\bibfield{author}{\bibinfo{person}{Oskar van Rest}, \bibinfo{person}{Sungpack
  Hong}, \bibinfo{person}{Jinha Kim}, \bibinfo{person}{Xuming Meng}, {and}
  \bibinfo{person}{Hassan Chafi}.} \bibinfo{year}{2016}\natexlab{}.
\newblock \showarticletitle{PGQL: a property graph query language}. In
  \bibinfo{booktitle}{\emph{Proc. 4th Int. Workshop on Graph Data Management
  Experiences and Systems}}. \bibinfo{pages}{7}.
\newblock


\bibitem[\protect\citeauthoryear{Wilschut and Apers}{Wilschut and
  Apers}{1991}]%
        {wilschut91}
\bibfield{author}{\bibinfo{person}{A.N. Wilschut} {and} \bibinfo{person}{P.M.G.
  Apers}.} \bibinfo{year}{1991}\natexlab{}.
\newblock \showarticletitle{Dataflow query execution in a parallel main-memory
  environment}. In \bibinfo{booktitle}{\emph{Proc. 1st Int. Conf. on Parallel
  and Distributed Information Systems}}. \bibinfo{pages}{68--77}.
\newblock


\bibitem[\protect\citeauthoryear{Wood}{Wood}{2012}]%
        {wood2012query}
\bibfield{author}{\bibinfo{person}{Peter~T Wood}.}
  \bibinfo{year}{2012}\natexlab{}.
\newblock \showarticletitle{Query languages for graph databases}.
\newblock \bibinfo{journal}{\emph{ACM SIGMOD Rec.}} \bibinfo{volume}{41},
  \bibinfo{number}{1} (\bibinfo{year}{2012}), \bibinfo{pages}{50--60}.
\newblock


\bibitem[\protect\citeauthoryear{Yakovets, Godfrey, and Gryz}{Yakovets
  et~al\mbox{.}}{2016}]%
        {yakovets2016query}
\bibfield{author}{\bibinfo{person}{Nikolay Yakovets}, \bibinfo{person}{Parke
  Godfrey}, {and} \bibinfo{person}{Jarek Gryz}.}
  \bibinfo{year}{2016}\natexlab{}.
\newblock \showarticletitle{Query planning for evaluating SPARQL property
  paths}. In \bibinfo{booktitle}{\emph{Proc. ACM SIGMOD Int. Conf. on
  Management of Data}}. \bibinfo{pages}{1875--1889}.
\newblock


\bibitem[\protect\citeauthoryear{Yang, Golab, and {\"O}zsu}{Yang
  et~al\mbox{.}}{2017}]%
        {yang:2017aa}
\bibfield{author}{\bibinfo{person}{Yuke Yang}, \bibinfo{person}{Lukasz Golab},
  {and} \bibinfo{person}{M.~Tamer {\"O}zsu}.} \bibinfo{year}{2017}\natexlab{}.
\newblock \showarticletitle{{ViewDF: D}eclarative Incremental View Maintenance
  for Streaming Data}.
\newblock \bibinfo{journal}{\emph{Inf. Syst.}}  \bibinfo{volume}{71}
  (\bibinfo{year}{2017}), \bibinfo{pages}{55--67}.
\newblock
\urldef\tempurl%
\url{https://doi.org/doi.org/10.1016/j.is.2017.07.002}
\showDOI{\tempurl}


\end{thebibliography}

\end{document}